%% file: HyperMRI_PNAS_arxiv.tex
\documentclass[aps,amssymb,amsmath,pra,twocolumn,showpacs,superscriptaddress]{revtex4-1}

\usepackage{graphicx}% Include figure files
\usepackage{dcolumn}% Align table columns on decimal point
\usepackage{bm}% bold math
\usepackage{subfigure}
\usepackage{color}

\usepackage{amssymb,amsmath,amsfonts,latexsym,graphicx,verbatim}%,dsfont}

\usepackage[margin=0.55in]{geometry}

\usepackage[english]{babel}
\usepackage{times}
\usepackage{latexsym}
\usepackage{fancyhdr}
\usepackage{float}
\usepackage{afterpage}
\usepackage{enumitem}
\usepackage{eso-pic, graphicx}

\usepackage{listings}
\usepackage{multirow}
\usepackage[table]{xcolor}

\usepackage{bbm}
\usepackage{upgreek}
\usepackage{amsmath}

\definecolor{Ablue}{rgb}{0.96,0.24,0.00}

\definecolor{Abluetitle}{rgb}{0.,0.24,0.51}
\newcommand{\bluetitle}{\color{Abluetitle}}

\definecolor{orange}{rgb}{0.96,0.24,0.00}

\definecolor{darkred}{rgb}{0.55, 0.0, 0.0}
\definecolor{lR}{rgb}{1, 0.8, 0.79}

\definecolor{Gray}{gray}{0.85}
\definecolor{lC}{rgb}{0.88,1,1}
\definecolor{darksalmon}{rgb}{0.91, 0.59, 0.48}
\definecolor{maroon}{cmyk}{0,0.87,0.68,0.32}

\definecolor{mustard}{rgb}{1.0, 0.86, 0.35}
\newcolumntype{a}{>{\columncolor{Gray}}c}
\newcolumntype{b}{>{\columncolor{white}}c}

\usepackage{array}
\newcolumntype{L}[1]{>{\raggedright\let\newline\\\arraybackslash\hspace{0pt}}m{#1}}
\newcolumntype{C}[1]{>{\centering\let\newline\\\arraybackslash\hspace{0pt}}m{#1}}
\newcolumntype{R}[1]{>{\raggedleft\let\newline\\\arraybackslash\hspace{0pt}}m{#1}}

\usepackage[colorlinks=true , citecolor=blue,urlcolor=blue]{hyperref}

\input{Commands3.tex}

\newcommand{\beginsupplement}{%
        \setcounter{table}{0}
        \renewcommand{\thetable}{S\arabic{table}}%
        \setcounter{figure}{0}
        \renewcommand{\thefigure}{S\arabic{figure}}%
				
     }

\usepackage{threeparttable}
\usepackage{booktabs,  makecell}
\newcommand{\RomanNumeralCaps}[1]
    {\MakeUppercase{\romannumeral #1}}
\newcommand{\affA}{Department of Chemistry, University of California, Berkeley, California 94720, USA.}
\newcommand{\affB}{Department of Physics and CUNY-Graduate Center, CUNY-City College of New York, New York, NY 10031, USA.}

\newcommand{\affD}{Department of Chemical and Biomolecular Engineering, and Materials Science Division Lawrence Berkeley National Laboratory University of California, Berkeley, California 94720, USA.}
\newcommand{\affE}{Department of Radiology and Biomedical Imaging, University of California—San Francisco, San Francisco, California 94158, USA.}

\newcommand{\affH}{Adámas Nanotechnologies, Inc., 8100 Brownleigh Dr, Suite 120, Raleigh, NC, 27617 USA.}
\newcommand{\affG}{Nuclear Magnetic Resonance Facility, University of California Davis, Davis, California 95616, USA.}

\begin{document}
\title{\bluetitle{Background-free dual-mode optical and $^{13}$C magnetic resonance imaging in diamond particles}}

   \author{X. Lv}\affiliation{\affA}
	 \author{J. H. Walton}\affiliation{\affG}
 \author{E. Druga}\affiliation{\affA}
\author{F. Wang}\affiliation{\affA}
\author{A. Aguilar}\affiliation{\affA}
\author{T. McKnelly}\affiliation{\affA}
\author{R. Nazaryan}\affiliation{\affA}
\author{L. Wu}\affiliation{\affA}
   \author{O. Shenderova}\affiliation{\affH}
	 \author{D. B. Vigneron}\affiliation{\affE}
	\author{C. A. Meriles}\affiliation{\affB}
   \author{J. A. Reimer}\affiliation{\affD}
   \author{A. Pines}\affiliation{\affA}
	\author{A. Ajoy}
	\email{ ashokaj@berkeley.edu}
	\affiliation{\affA}
	
\begin{abstract}
Multimodal imaging -- the ability to acquire images of an object through more than one imaging mode simultaneously -- has opened new perspectives in areas ranging from astronomy to medicine. In this paper, we report progress towards combining optical and magnetic resonance (MR) imaging in such a ``dual'' imaging mode. They are attractive in combination because they offer complementary advantages of resolution and speed, especially in the context of imaging in scattering environments. Our approach relies on a specific material platform, microdiamond particles hosting nitrogen vacancy (NV) defect centers that fluoresce brightly under optical excitation and simultaneously ``hyperpolarize" lattice $\Cs$ nuclei, making them bright under MR imaging. We highlight advantages of dual-mode optical and MR imaging in allowing background-free particle imaging, and describe regimes in which either mode can enhance the other. Leveraging the fact that the two imaging modes proceed in Fourier-reciprocal domains (real and k-space), we propose a sampling protocol that accelerates image reconstruction in sparse-imaging scenarios. Our work suggests interesting possibilities for the simultaneous optical and low-field MR imaging of targeted diamond nanoparticles.

\end{abstract}

\maketitle

In the quest towards high signal-to-noise (SN) imaging, significant power can be brought to bear by \I{multimodal} or \I{multi-messenger} techniques~\cite{lee12,abbott17}. They entail capturing an object through more than one imaging mode simultaneously, often at widely disparate wavelengths. Exploiting correlations between the different modes portend approaches such as Kalman filtering~\cite{welch1995introduction} that can deliver noise or background suppression. Furthermore, under appropriate conditions these correlations can engender novel image sampling and reconstruction strategies to accelerate image acquisition.

In this paper, we consider whether such methods could be applied to combine together MRI and optical imaging~\cite{tovar13,oto13}. These two modes offer diametrically complementary advantages. Visible-wavelength optics is fast, cheap and can image at high-resolution; yet often suffers from scattering, attenuation, and aberration distortions while imaging through real media. MRI, on the other hand, is noninvasive and scattering-free, fully three-dimensional and can be chemically functional; yet it is slow, suffers from weak signals, and offers poor spatial resolution (mm-level). Notably, optical and MR imaging are carried out in Fourier-reciprocal spaces (x- and k-space). This redundancy makes a combined modality attractive -- not only are there complementary advantages in sensitivity and resolution to be gained, but it opens possibilities to access hybrid acquisition strategies that sample both real and k-space \I{simultaneously} to yield image acceleration. 

Here we illustrate proof-of-concept demonstrations of dual-mode optical and $\Cs$ MR imaging in diamond microparticles. We demonstrate high fidelity imaging in either mode, and show they can be rendered background-free. We propose three imaging regimes wherein such dual-mode imaging provides benefits over either mode considered individually. In particular, in either scattering and scattering-free media, optics and MRI can complement each other with respect to resolution, SNR and depth of imaging. We propose a hybrid sampling strategy, wherein conjugate imaging is carried out simultaneously in real and k-space to enable image acceleration and power reduction in wide field-of-view settings.

Combined optical-MR imaging is made possible by special features of the diamond material medium. The diamond particles are incorporated with  $\gtrsim$1ppm of Nitrogen Vacancy (NV) defect centers~\cite{Jelezko06}. Under sub-bandgap illumination $<$575 nm, the particles fluoresce brightly in the red with high luminosity ($\sim$90cd/m$^2$) and optical stability. Fluorescence occurs {concurrently} with the optical polarization ($>$10\%) of the \NV electronic spins. This can be transferred to $\Cs$ nuclei in the surrounding lattice, {hyperpolarizing} them, and making them amenable to direct MR imaging~\cite{Fischer13}. We exploit a recent \NV-mediated hyperpolarization technique~\cite{Ajoy17} that allows large ($\sim$1\%) $\Cs$ polarization levels in diamond particles at room-temperature and low-magnetic fields.

\T{\I{Results}} -- \zfr{dual_image}A is a schematic of the experiment. Diamond particles (200$\mu$m size, $\sim$40mg) arranged in a ring-shaped phantom (\zfr{dual_image}D) are imaged optically under continuous 520nm illumination and 630nm long-pass filtering (\zfr{dual_image}E). The particles fluoresce brightly ($\sim$12$\zt 10^{11}$cps~\cite{SOM}). The same optical excitation polarizes (initializes to $m_s$=0) the \NV electron spins, and microwave (MW) sweeps across the \NV ESR spectrum drive Landau-Zener (LZ) dynamics that transfers polarization to the $\Cs$ nuclei in an orientation-independent manner (\zfr{dual_image}B)~\cite{Ajoy17,Zangara18}. We obtain $\sim$0.3\% $\Cs$ spin polarization in 40s under 1W total optical illumination enabling hyperpolarized MR imaging (\zfr{dual_image}B). This level corresponds to a signal enhancement of 280 times over thermal $\Cs$ polarization at 7T (\zfr{dual_image}C) or 206 times over 9.4T ($\sim$5$\zt$10$^4$ over 38mT), and $\sim$10$^5$ acceleration in MR imaging time.

Compared to conventional Dynamic Nuclear Polarization (DNP) techniques, our method requires relatively low laser ($\sim$2mW/mg) and MW power ($\sim$0.05mW/mg)~\cite{ajoy2020room}. The MRI demonstration in \zfr{dual_image}F employed a laser power density of $\sim$80mW/mm$^2$ to polarize $\sim$40mg diamonds. While this work is focussed on diamond particle imaging, we note possible extension for \textit{in-vivo} studies assuming that SAR can be controlled to safe limits, and the diamond particles can be eliminated from the body. In the context of potential in-vivo applicability, we estimate a MW specific absorption rate (SAR) of 1.1$\zt$10$^4$ W/kg[G]$^2$. Hyperpolarization efficiency scales approximately logarithmically with MW power beyond a particular threshold~\cite{Ajoy17}, indicating that the SAR can potentially be curtailed without severely degrading the DNP enhancement factors (see ~\cite{SOM}).

We use a variant of FLASH~\cite{Haase90} to produce the MRI images at 9.4T (see \zfr{dual_image}B, F), with short echo times (0.5ms) to accommodate short $T_2\app$1ms of $\Cs$ in diamond. To eliminate pulse interference during the $\sim$200$\mu$s gradient switching periods, imaging was performed without a slice selection gradient (\zfr{dual_image}B). The SNR of the MR image (\zfr{dual_image}F) is $\sim$4 in 16 scans (each scan preceded by hyperpolarization), limited by rapid $\Cs$ $T_2$ decay, low sample filling factor ($\app$0.007) and laser-limited hyperpolarization. The use of dynamic decoupling sequences, such as quadratic echos~\cite{Frey12}, spin-locking~\cite{Rhim76} or a recently developed approach to enhance FID time exceeding 2s~\cite{ajoy2020dynamical}, can improve the imaging SNR by at least an order of magnitude. If the application at hand permits higher optical powers close to saturation intensity, similar gains can be concurrently obtained. In particular, materials advances would boost MR image SNR by an order of magnitude through $\Cs$ enrichment, and through the use of high temperature annealing~\cite{Gierth2020}.
Further improvements may be realized by optimizing the detection coil geometry and filling factor, for instance through the use of small volume inductively-coupled receiver coils. These concerted gains in MR signal could also permit similar high-contrast images in nanosized particles ($<$100nm), although these smaller particles display lower ($\app 10^{-2}$) hyperpolarizability than the microparticles employed in this work~\cite{Gierth2020}. The MRI spatial resolution here is 640$\mu$m $\propto 1/(\gamma G_{\R{max}}\tau)$, where $G_{\R{max}}$ = 950mT/m. 

The DNP method here presents advantages over traditional hyperpolarization methods for solids imaging, employed for instance in $^{29}$Si microparticles~\cite{Cassidy2013}. We work at room temperature and low field ($\sim$40 mT), and polarize samples in under 1min of laser pumping. Conventional methods, in contrast, require high magnetic field ($\gtrsim$3T) and low temperature ($<$4K) and polarization buildup can take several hours~\cite{Bretschneider16}. While the absolute polarization is lower in our method, we circumvent the high polarization loss (as large as 99\%~\cite{Waddington19}) accrued upon thawing and sample transfer out of the cryostat. Technologically, our technique aids end-user operation -- MW amplifiers and sweep sources are low-cost and readily available, and hyperpolarized particles can be delivered by a portable device~\cite{ajoy2020room}. Since the DNP process is detection-field agnostic, the technique is especially interesting in the context of low-field MRI where hyperpolarization can be replenished continuously.

\begin{figure*}[t]
  \centering
  {\includegraphics[width=0.85\textwidth]{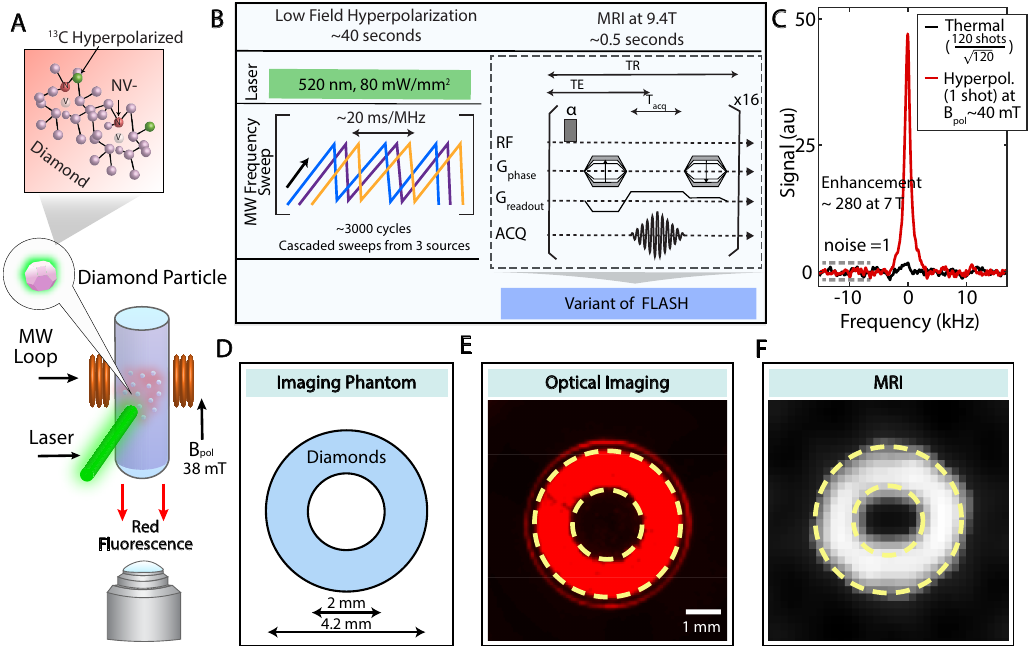}}
  \caption{\T{{Dual-mode optical and $\Cs$ MR imaging.}} (A) \I{Experiment schematic.} Diamond particles with \NV centers are imaged with fluorescence under green (520nm) excitation by a CMOS detector, as well as under $\Cs$ MRI through polarization transferred to lattice $\Cs$ nuclei from optically polarized \NV electrons. (B) \I{Hyperpolarization and detection protocol.} $\Cs$ DNP occurs at low-field $\sim$38mT under MW sweeps across the \NV ESR spectrum. FLASH MR imaging is performed after sample shuttling to 9.4T. Here echo time TE = 0.5ms, repetition time TR = 6ms, acquisition time T$_{\R{acq}}$ = 0.36ms. (C) \I{Typical hyperpolarization signal enhancement}, showing signal gain $\app$280$\pm$5 against thermal signal at 7T, corresponding to $\sim$5 orders of magnitude acceleration in MR imaging time. For a fair comparison, the noise in both datasets is normalized to 1 (dashed lines). (D) Ring shaped phantom filled with 40mg 200$\mu$m diamond particles employed for dual-mode imaging. (E) \I{Fluorescence image} captured through a 630nm long-pass filter. (F) \I{$\Cs$ MR FLASH image} with 16 averages ($\app$8s total imaging time), and a square pixel length of 160$\mu$m and square FOV with 6.4mm edge.}
\zfl{dual_image}
\end{figure*}

\begin{figure*}[t]
  \centering
  {\includegraphics[width=0.9\textwidth]{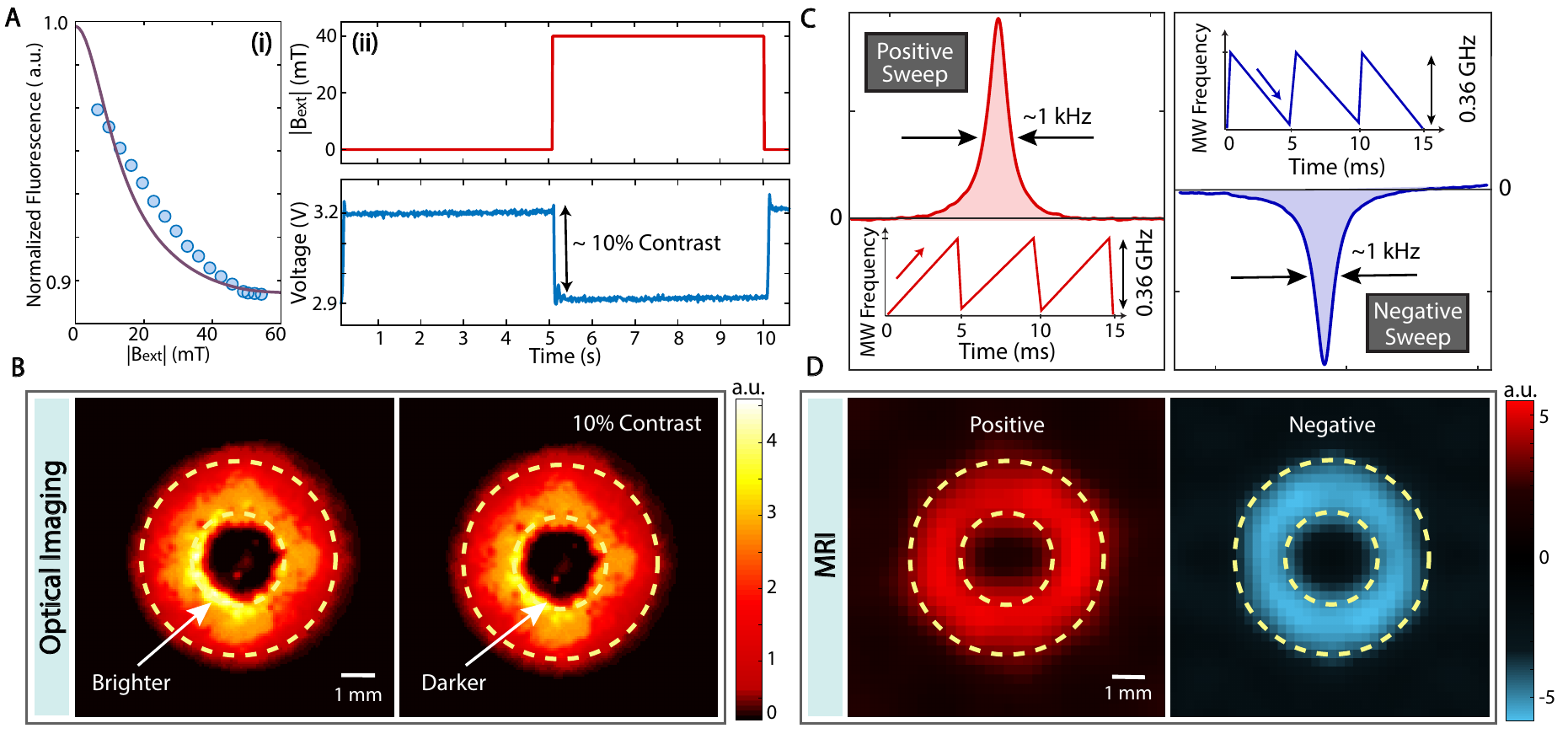}}
  \caption{\T{{On-demand dual-mode image modulation.}} (A) (i) Normalized fluorescence signal for randomly oriented diamond particle ensemble under an applied magnetic field (points: experiment, purple-line: simulation). We ascribe the discrepancy to scattering effects. (ii) Optical modulation under 40$\pm$2mT pulsed magnetic field showing a signal contrast $\sim$10\%. (B) \I{Optical images} under 0 and $\sim$40mT applied field showing weak $\sim$10\% optical contrast. (C) \I{$\Cs$ hyperpolarization sign control}. MW frequency sweeps in low-to-high (high-to-low) fashion across the \NV ESR spectrum leads to positive (negative) hyperpolarization. Shown are 7T $\Cs$ NMR spectra under opposite sweep conditions. (D) \I{$\Cs$ MR images} under opposite MW sweep conditions, showing full sign reversal and $\sim$194\% modulation contrast. Here FLASH images were taken with TE=0.6ms and TR=6ms.}
\zfl{pn_images}
\end{figure*}

\T{\I{Background-free imaging}} -- Both optics and MR modalities allow on-demand image amplitude modulation, enabling common-mode suppression of background signals. The NV fluorescence is conditioned strongly on the misalignment angle $\xt$ of the N-to-V axis to the applied field~\cite{tetienne12,Liu19}, especially at low fields approaching 50mT. This arises from mixing of the $m_s=\pm$1 spin levels in the excited state. Since the randomly oriented particles sample all possible $\xt$ angles, this enables a simple method to modulate the optical images by applying a pulsed field ${\mathbf{B}_{ext}}$~\cite{bumb14}. In \zfr{pn_images}A (i),  we simulate fluorescence dependence under ${\mathbf{B}_{ext}}$ using a 7-level model of the NV center~\cite{tetienne12}:  $\frac{\mathrm{d} n_{i}}{\mathrm{d} t}=\sum_{j=1}^{7}\left(k_{j i} n_{j}-k_{i j} n_{i}\right)$, where $n_i$ is the population of the $\ket{i}$ state, and $k_{ij}$ denotes the kinetic transition rate between states $\ket{i}$ and $\ket{j}$ (measured in \cite{tetienne12}). In steady state, photoluminescence  (purple line in \zfr{pn_images}A (i)), obtained evaluating $\int n_i(\mathbf{B}_{ext},\xt)\sin\xt d\xt$) decreases with ${\mathbf{B}_{ext}}$, in reasonable agreement with (normalized)  measurements (blue dots). In our experiments (\zfr{pn_images}B), ${\mathbf{B}_{ext}}=$40 $\pm$2mT takes the value identical to that used for hyperpolarization, and we obtain a $\sim$10\% optical modulation contrast. (\zfr{pn_images}A (ii)). 

MRI mode allows for similar signal modulation. Such modulation refers to controlling the $\Cs$ hyperpolarization \I{sign} based on the direction of MW sweeps between alternate scans (as opposed to continuously modulated signal in optics). This effect originates from LZ dynamics excited by the chirped microwaves (\zfr{pn_images}C)~\cite{Ajoy17,zangara2019dynamics}. The $\Cs$ nuclei are aligned (anti-aligned) with the polarization field under low-to-high (high-to-low) frequency sweeps. This allows {complete} sign-reversal of the MRI images at full contrast.   As a figure of merit, we characterize modulation contrast as the difference ratio of the MR images $\mathcal{I}$ and $\overline{\mathcal{I}}$ under opposite MW sweeps (\zfr{pn_images}D) as: $\xD = \frac{1}{N^2}\lb\frac{\mathcal{I}- \ov{\mathcal{I}}}{\mathcal{I}+\ov{\mathcal{I}}}\rb$, where $N^2$ is the total number of pixels. From the data in \zfr{pn_images}E-F, we obtain $\xD=($194 $\pm$ 0.3)\%. Similar modulation contrasts are challenging to achieve in conventional cryogenic DNP due to technical limitations of MW cavity switching~\cite{Becerra93,Waddington19}.

Signal modulation allows background-free imaging of the diamond particles. We refer to "{background}'' in this context as media with fluorescence \I{or} $\Cs$ NMR signals that overlap in wavelength (or NMR frequency) with the diamond particles.   As a proof of concept (\zfr{bg_spp}), we consider particles being co-included with a high concentration of Alexa 647, a fluorescent dye with emission in the 650-670nm range~\cite{SOM}, as well as [$^{13}$C]-methanol, which has a chemical shift nearly overlapping that of diamond. These solution media fill both the inner and outer spaces of the capillary tube that comprise the diamond phantom in \zfr{bg_spp}A-B. The backgrounds result in images that are circle-shaped since the diamond phantom is completely indiscernible within it (\zfr{bg_spp}C-D). To recover the diamond signals in the optical mode, we perform lock-in detection under a 40mT 0.1Hz square-modulated field. We record a 2000-frame movie at 0.1fps, and computationally apply lock-in suppression on each pixel. The resulting image (\zfr{bg_spp}E) shows recovery of the diamond signal, in this case, from the 2 times stronger background. Optical background suppression is ultimately limited by the modulation contrast, and up to an order of magnitude is experimentally feasible~\cite{bumb14}. Concurrent MRI background suppression is realized by subtracting the images under opposite sweep-ramp hyperpolarization conditions in alternative scans. The [$^{13}$C]-methanol signal was 5 times larger than the diamond signal, and is efficiently canceled (\zfr{bg_spp}F). 

\begin{figure}[t]
  \centering
  {\includegraphics[width=0.49\textwidth]{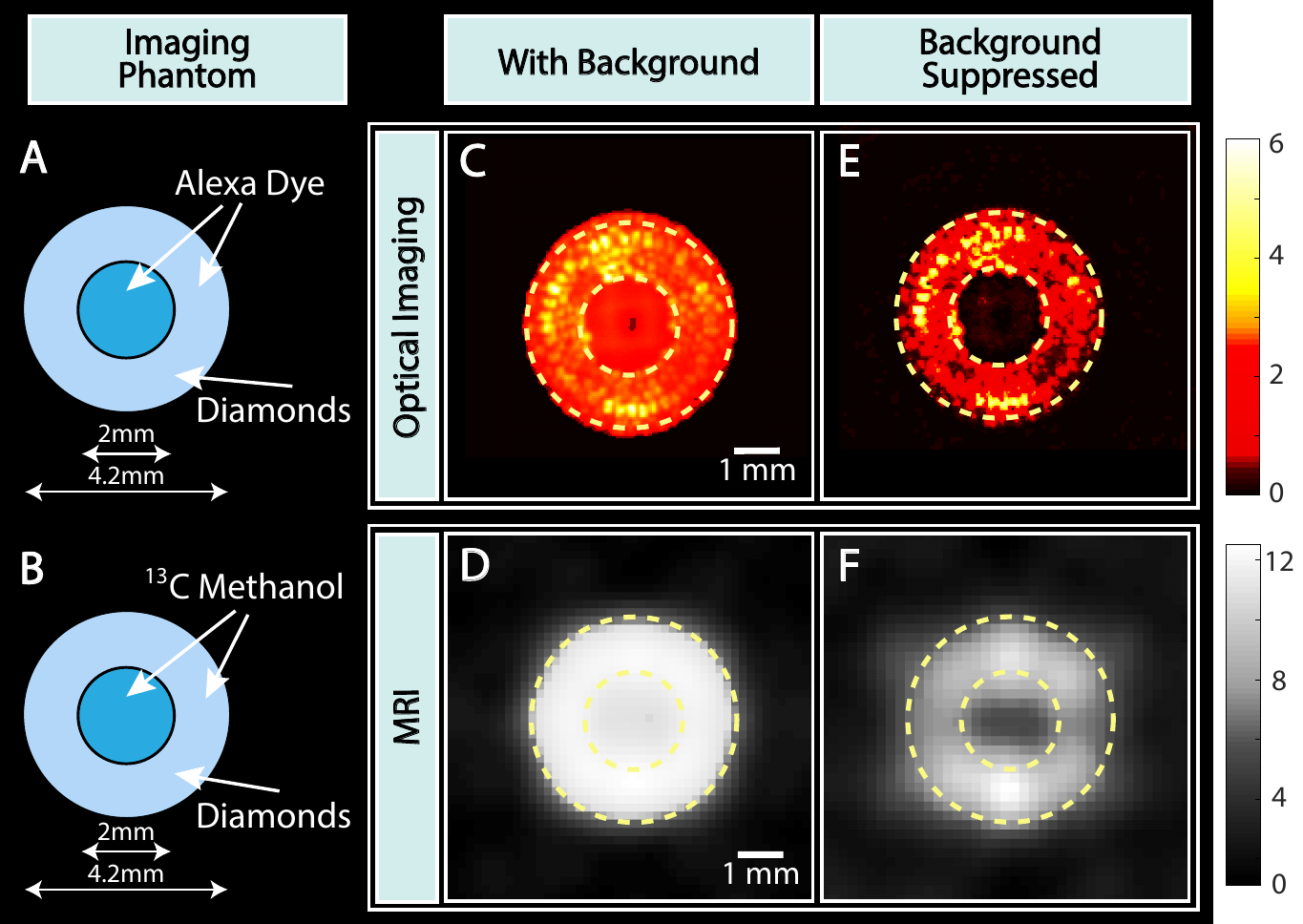}}
  \caption{\T{Dual-mode background suppression.} (A-B) \I{Schematic of imaging phantoms}. Diamonds are arranged in ring-shaped phantom, {co-situated} with Alexa 647 dye and $^{13}$C-methanol that present an artificial backgrounds for optical and MR imaging respectively. (C-D) Optical and MR images with the background. Dashed lines serve as a guide to the eye for the imaging phantom. Diamond particles are indistinguishable from the background in both imaging dimensions. (E-F) \I{Background suppressed} optical and MR images employing signal modulation (reversal) allows complete recovery of the original diamond phantom in both imaging modes.}
\zfl{bg_spp}
\end{figure}

\begin{figure*}[t]
  \centering
  {\includegraphics[width=0.9\textwidth]{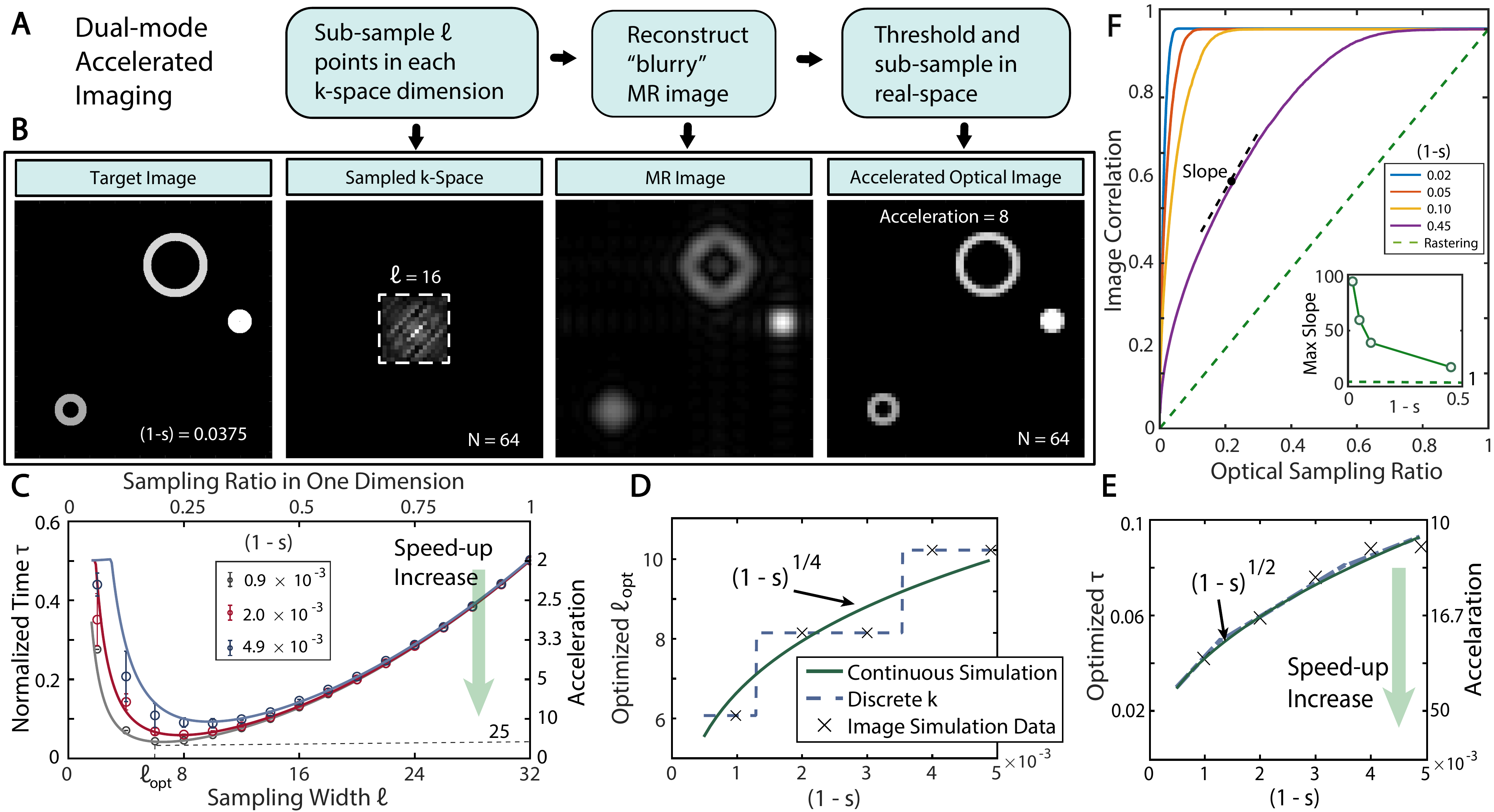}}
\caption{\textbf{Accelerated \I{x-k} conjugate-space imaging.} (A) \I{Protocol for accelerated imaging.} $\ell$ samples of the image are first acquired in k-space, and the resulting image upon thresholding is fed-forward to constrain the real-space points to be scanned over. (B) \I{Exemplary scenario} where the target image FOV consists of sparsely distributed objects. By sampling over $\ell$=16 k-space points in each dimension, a blurry yet faithful image is formed of the target (center two panels), and can serve to restrict real-space sampling (right panel), leading to $\sim$ 14 times acceleration. (C) \I{Normalized imaging time} $\qt$ with different values of $\ell$ for images in a 32$^2$ pixel square FOV of different sparsity factors. Right axis indicates corresponding imaging acceleration. Here images consisted of unit-pixel objects, and we averaged over 30 random image configurations with identical sparsity; error bars denote standard deviations. The presence of an optimal k-space sample threshold $\ell_{\R{opt}}$ is evident, stemming from a compromise between better confinement in real-space imaging, and associated time-cost for k-space imaging. (D) \I{Scaling of $\ell_{\R{opt}}$} with image sparsity, showing that more k-space values are required to account for increasing imaging complexity. Points: given discrete possible values of $k$, $\ell_{\R{opt}}$ has a staircase-like behavior. Solid line: scaling of $\ell_{\R{opt}}$ assuming continuous $k$ values. (E) \I{Optimized imaging acceleration} through sampling $\ell_{\R{opt}}$ points, showing orders of magnitude time savings at high sparsity. (F) \I{Trajectory of image convergence} quantified by image correlation ${\cal C}$ with protocol advancement. Green dashed line shows linear convergence under conventional optical rastering. In contrast, hybrid sampling in conjugate spaces can lead to rapid image convergence. \I{Inset:} Slope of image approach to target rapidly increases with sparsity.}
\zfl{fast_image}
\end{figure*}

\T{\I{Regimes for combined dual-mode imaging}} -- While \zfr{bg_spp} shows the optical and MR imaging separately, what may one accomplish by \I{combining} them together in a dual-imaging modality? Consider that MRI is natively immune to optical scattering, while optics provides superior resolution in scattering-free media. This can make a combined mode persuasive in specific imaging regimes. 

Consider first the fluorescent imaging of nanodiamond particles embedded in a scattering medium (eg. in tissue). One has to contend with \I{round-trip} attenuation and scattering losses (scaling $\sim\xl^{-4}$) that restricts the proportion of collected photons. In tissue, for instance, attenuation and scattering at 650nm lead to {exponential} losses with coefficients 0.1mm$^{-1}$ and 1.1mm$^{-1}$, respectively~\cite{Jacques96,Lister12}. We estimate a round-trip attenuation loss of 0.2 and scattering loss of $10^{-4}$ at 3mm depth. There are further photon losses stemming from green-red-conversion ($\sim 10^{-5}$), the unusually high refractive index differential between diamond and its environment ($\sim 10^{-2}$), as well as finite numerical aperture limited by geometric solid angle constraints of the detector at high FOVs ($\sim 10^{-3}$) (see \cite{SOM}). Scattering also leads to a simultaneous loss in imaging resolution. Random scattering of the beam blurs the optical image, spreading it as much as 800$\mu$m at 1cm depth (see~\cite{SOM} for simulations).

In contrast, hyperpolarized MR imaging SNR can be competitively efficient, especially at increasing imaging depth $d$.  One only partakes of losses in the \I{one-way} illumination of the particles with 520nm light. Hyperpolarization efficiency is $\sim$2$\zt$10$^{-4}$ per $\Cs$ nucleus per incident photon~\cite{Ajoy17}. The high detection losses and geometric solid-angle collection constraints are replaced by more benign factors related to sample-coil filling, detector Q, and the overall MR detection frequency~\cite{Hoult1978}. Surface coils matched to the sample and the use of high-Q ferrite resonators can lead to substantially more efficient detection~\cite{Ackerman80}. In this regime MR imaging could have a higher overall SNR than its optical counterpart. This crossover point in efficiency occurs at depths $d\app\fr{1}{1.2}\log(\eta_d)$ at 650nm, where $\eta_d$ is the ratio of optical and MR imaging SNR for \I{surface} diamond particles. Immunity to scattering also means that MR imaging resolution is independent of imaging depth~\cite{SOM}.

To now elucidate advantages of dual-mode optical-MR imaging, we consider the effect of one mode enhancing the other. We focus on two imaging regimes, conditioned on imaging in media with and without optical scattering respectively. In each regime we consider a ``primary" imaging mode (either optics or MRI) and wherein a ``secondary" (complementary) mode is added to enhance it (see Table \ref{3regime_table1}). 

\I{Regime I:} Consider first optical imaging (primary mode) in scattering media. Scattering deteriorates image SNR by a factor of $e^{-\alpha_i d}e^{-\alpha_\epsilon d}$, where $\alpha_i$, $\alpha_\epsilon$ are incident and fluorescent loss coefficients respectively. Instead, the ability to image via the secondary MRI mode can improve SNR compared to the primary mode by a factor $e^{-\alpha_\epsilon d}$, and imaging resolution ($\propto d$). Table \ref{3regime_table1} shows these factors, with estimated numerical gains possible in the scenario with imaging depth $d=$3mm (see footnotes in Table \ref{3regime_table1}). In fact, the depth penetrability of MRI, and its relative immunity to scattering can also provide imaging capability in 3D.

\I{Regime II:} Alternately, consider when MRI is chosen as the primary mode and is applied in a scattering-free imaging regime. Here the resolution can be augmented from $\delta x_{\mathrm{mri}}$ (640 $\mu$m in \zfr{dual_image}) to $\delta x_{\mathrm{optics}}$ (40 $\mu$m in \zfr{dual_image}, ultimately limited by diffraction) by rastering a laser beam across the field of view to selectively hyperpolarize one single pixel at a time. Assuming the available total power is fixed, focusing the beam increases power density by a factor of $\frac{\mathrm{FOV}}{\Delta x_{\mathrm{optics}}^2}$. Despite pixel size change and imaging time extension, these factors can provide an SNR enhancement of $\frac{\sqrt{FOV}\Delta x_{\mathrm{optics}}}{\Delta x_{\mathrm{mri}}^2}$ simultaneously with resolution gain.

\T{\I{Accelerated conjugate-space imaging}} -- While the discussion above considers dual-mode imaging in terms of relative merits of optics and MRI with respect to scattering and resolution, here we propose an additional regime (\I{Regime III}) exploiting the native ability of the two imaging methods to sample in Fourier reciprocal spaces. This allows feed-forwarding information from one space to guide the sampling in the other space. Every sampled point in one space carries information from \I{all} the points in the other space; this is exploited to provide speedup in sparse imaging settings. This approach shares similarities with compressed sensing~\cite{Donoho06,Lustig07} (see \cite{SOM} for a detailed comparison), but is different in the sense that image sampling here physically occurs in two conjugate spaces.

 As a specific example, consider a wide field-of-view (FOV) imaging scenario where optical imaging is performed by rastering a low-power beam across a sample of nanodiamonds distributed in a volume. We assume the objects are sparse in FOV and, for simplicity, that the per-pixel imaging time cost is identical for both optics and MRI. \zfr{fast_image} describes the imaging protocol. Given the sparse original image to be acquired, a subset of $\ell$ points in each dimension are first sampled in k-space via MRI. The resulting blurry low-k image is thresholded and {fed-forward} to confine real-space points to be sampled through optics. At high sparsity $s$, defined herein as the fraction of zero pixels in the FOV, this k-space reduces the number of points in real-space to be rastered over, and accelerates image acquisition. \zfr{fast_image}B simulates the method applied to an image with high sparsity, $(1-s)=$0.0375. Employing $\ell$=16 k-space samples in each dimension, and thresholding at 10\% of the mean, we estimate image acceleration of 16-fold by eliminating sampling from large parts of the (real-space) image.

Let us now analyze the regime of applicability and inherent trade-offs involved. For simplicity, we assume as before that the time cost to be accrued per sample (pixel) is identical for both optical and MR imaging dimensions, although it is straightforward to scale the results by the cost-ratio $\eta$ as appropriate. There is an \I{optimal} k-space sample $\ell_{\R{opt}}$, arising as a compromise between better constraining real-space sampling, and taking longer to image. \zfr{fast_image}C demonstrates this for a 32 $\times$ 32 pixel FOV, where we consider normalized imaging time savings over either modality for target images with varying sparsity to identify $\ell_{\R{opt}}$. We estimate that hybrid sampling can deliver more than an order of magnitude in time savings (right axis in \zfr{fast_image}C ), while only requiring the scanning of $\ell_{\R{opt}}\app$10\% of total k-space (upper axis). As expected, $\ell_{\R{opt}}$ decreases with sparsity, a reflection that larger k-samples are required with increasing image complexity (see \zfr{fast_image}D). Given the small FOV and discrete values of k-samples, this manifests in the staircase-like pattern in \zfr{fast_image}D, but scales $\ell_{\R{opt}}\propto (1-s)^{1/4}$ (solid line) as we shall derive below. Finally, in \zfr{fast_image}E we consider the combined imaging time $\qt$ under optimized conditions as a function of image sparsity, assuming that time for optical imaging is 1. Indeed, the imaging acceleration can be quite substantial, scaling as $\qt^{-1} \propto (1-s)^{-1/2}$.

We now elucidate origins of the imaging acceleration by studying the \I{convergence trajectory} of the reconstructed image as it approaches the target with each step of the protocol (see \zfr{fast_image}F). Considering several image configurations with a fixed sparsity, we analyze in \zfr{fast_image}F the overlap of the reconstructed image $\mI'$ to the target image $\mI$ through the correlation ${\cal C}=\sum\left(\mI-\expec{\mI}\right)\left(\mI'-\expec{\mI'}\right)$, where $\expec{\cdot}$ indicates the mean value. Indeed under usual rastered optical image acquisition (dashed green line in \zfr{fast_image}F), the reconstructed image {linearly} approaches the target as more samples are acquired. In contrast, employing the hybrid acquisition of a few k-space points, and by constraining the space over which the final image is to be acquired, one obtains a {rapid} convergence with the target. Numerically, the {slope} of convergence scales approximately $\propto (1-s)^{-1/2}$ (inset in \zfr{fast_image}F), indicating rapid gains can be amassed at high image sparsity.
To analytically elucidate image acceleration at high sparsity (solid lines in \zfr{fast_image}D-E), let us consider the real-space target image $f(x, y)$, which in k-space is $\mathcal{F}[f] =\hat{f}(k_x, k_y)$, where $\mF$ denotes a spatial Fourier transform. As k-space sampling now occurs just to $\ell$-th order, one obtains the reconstructed image, $\hat{f}_{\ell}(k_x, k_y) = \hat{f}(k_x, k_y) \cdot \Pi \left( \frac{k_x}{W_{kx}} \right) \cdot \Pi \left( \frac{k_y}{W_{ky}} \right)$, where $\Pi$ is a rectangular function representing a sampling window with a side length of, for instance, $W_{kx} = \delta k_x\ell$, where $\delta k_x=1/N_x$ is the k-space pixel size. Transformation back to real-space gives the convolution, $\mF^{-1}[\hat{f_{\ell}}] = f(x,y) * (W_{kx} \sinc (W_{kx} x) \cdot W_{ky} \sinc (W_{ky} y))$. An object of pixel radius $r$ is effectively blurred $r\rt r_0\lb\frac{N}{\ell}\rb$, where the factor $r_0$ is set by the thresholding level employed (see Supplementary Information~\cite{SOM}), and $\ell/N$ is the effective k-space sampling ratio. Increasing $\ell$ makes a more faithful representation and improves regional constraints, but it is associated with a time-cost. To analytically evaluate the time savings let us as an example consider the FOV consists of $n_d$ objects of radius $r$, giving $(1-s) = n_d\pi r^2/N^2$. The normalized imaging time is then, $\tau = (1-s)+\frac{2r_0}{r}\frac{1}{\ell} \left[ (1-s)N \right]+\frac{r_0^2}{r^2}\frac{1}{\ell^2}\left[ (1-s)N^2 \right]+\frac{\ell^2}{N^2} $~\cite{SOM}. In the limit of high sparsity, $(1-s)\rt 0$, $(1-s)N\rt 0$, and $(1-s)N^2\rt {\cal O}(1)$, giving $\tau\app  C_0\frac{(1-s)}{\ell^2} + \fr{1}{N}\ell^2$, where $C_0 = \frac{r_0^2N^2}{r^2}$ is a constant. Determining the optimal $\ell$ to minimize $\qt$, gives, $\ell_{\R{opt}} \propto (1-s)^{1/4}$, and the optimal (normalized) time $\qt\propto (1-s)^{1/2}$. The scaling from this simple model is shown as the solid lines in \zfr{fast_image}D-E and closely match numerical results in the limit of high sparsity. 

Finally, we comment that imaging acceleration results in a lower total optical power delivered to the sample by the same factor as the acceleration gain. This protocol might have real-world applications given that most imaging of targeted nanodiamond operates in the high-sparsity limit $s>$95\%~\cite{Schrand09,Mochalin12}.

\begin{table*}[!ht]
\begin{threeparttable}

\begin{tabular}{ |L{3.9cm}||L{4.2cm}|L{4cm}|L{4cm}|   }
 \hline
 %\multicolumn{4}{|c|}{Three imaging regimes} \\
 %\hline
Particular specifics	& \multicolumn{1}{|c|}{Regime \RomanNumeralCaps{1} }& \multicolumn{1}{|c|}{Regime \RomanNumeralCaps{2}} & \multicolumn{1}{|c|}{Regime \RomanNumeralCaps{3}}\\
	\hline
   Primary mode  & Optical imaging & Hyperpolarized MRI &  Accelerated dual-mode  \\
   \cline{1-3}
   Secondary mode & Hyperpolarized MRI & Optical imaging & imaging protocol \\
 \hline
Operating regime   &  \cellcolor{lR} Deep in scattering media & \cellcolor{lC}Shallow or in non-scattering media & \cellcolor{lC}Shallow or in non-scattering media\\
 \hline
Primary mode SNR &  \cellcolor{lC} $\propto \gamma_0 \cdot e^{-\alpha_i d}e^{-\alpha_\epsilon d}$ & \cellcolor{lC} $\propto {P_{i}} (\Delta x_{\mathrm{mri}})^2\cdot \sqrt{T}$ &  \cellcolor{lR} \\

Dual mode SNR & \cellcolor{lC} $ \propto e^{-\alpha_i d}$ & \cellcolor{lR} $ \propto \frac{\mathrm{FOV}}{\Delta x_{\mathrm{optics}}^2}  {P_{i}} (\Delta x_{\mathrm{optics}})^2\cdot \sqrt{T/\frac{\mathrm{FOV}}{\Delta x_{\mathrm{optics}}^2}}$ &  \multirow{-2}{*}{$ \propto \sqrt{T/(1-s)^{1/2}}$ \tnote{\emph{c}} \cellcolor{lR} }\\

SNR gain & $1/\gamma_0 \cdot e^{\alpha_\epsilon d}$ & $\frac{\sqrt{FOV}\Delta x_{\mathrm{optics}}}{\Delta x_{\mathrm{mri}}^2}$ &    $\propto 1/(1-s)^{1/4} $ \tnote{\emph{g}} \\
Primary mode SNR (example)  & \cellcolor{lC}4.4$\times10^{-3}\sqrt{\mathrm{Hz}}  $  \tnote{\emph{b}} & \cellcolor{lC} 0.63$\sqrt{\mathrm{Hz}}$ \tnote{\emph{d}} &  \cellcolor{lR} \\ 
Dual mode SNR (example) & \cellcolor{lC}1.1$\times10^{-2}\sqrt{\mathrm{Hz}}  $ \tnote{\emph{b}} &  \cellcolor{lC} 6.3$\sqrt{\mathrm{Hz}}$ \tnote{\emph{e}} &  \multirow{-2}{*}{\cellcolor{lR}47.5$\mathrm{\sqrt{Hz}}$ \tnote{\emph{d}} }\\
SNR gain (example)& \cellcolor{lC}2.5 & \cellcolor{lR} 10 & \cellcolor{lC}5 \tnote{\emph{g}} \\
 \hline
Primary mode resolution $\delta x_p$ & $ \propto d$ & $\delta x_{\mathrm{mri}}$ & \\
Dual mode resolution $\delta x_d$  & $\sim \frac{1}{\gamma G_{max} \tau}$ & $\delta x_{\mathrm{optics}}$ &\multirow{-2}{*}{ $\delta x_{\mathrm{optics}}$}\\
Resolution gain $\frac{\delta x_p}{\delta x_d}$ & $ \propto d$  & $\frac{\delta x_{\mathrm{mri}}}{\delta x_{\mathrm{optics}}}$ & N.A.\\
Resolution limit &  $\sim \frac{1}{\gamma G_{max} \tau}$ ($\sim$ 1$\mu$m) \tnote{\emph{h}}& $ \frac{\lambda}{2} $ & $ \frac{\lambda}{2} $\\
Primary mode resolution (example) & \cellcolor{lC}1200$ \mu$m \tnote{\emph{c}} & \cellcolor{lC} 640$\mu$m \tnote{\emph{d}} & \cellcolor{lR}  \\ 
Dual mode resolution (example) & \cellcolor{lC} 640$\mu$m \tnote{\emph{d}} &  \cellcolor{lR} 40$\mu$m \tnote{\emph{d}} & \multirow{-2}{*}{\cellcolor{lR} 40$\mu$m \tnote{\emph{d}}}\\
Resolution gain (example)& \cellcolor{lC}  1.875 &  \cellcolor{lR} 16 &  N.A.\\
 \hline
Power reduction \tnote{\emph{a}} & $1/\gamma_0 \cdot e^{\alpha_\epsilon d}$ & $\frac{\sqrt{\mathrm{FOV}}\Delta x_{\mathrm{optics}}}{\Delta x_{\mathrm{mri}}^2}$ & $\propto 1/(1-s)^{1/2}$\\
Power reduction (example) & \cellcolor{lC} 2.5 & \cellcolor{lC} 10 &  \cellcolor{lR} 25 \tnote{\emph{f}} \\
\hline
Background suppression & MRI can suppress optical background & Minimal background in $\Cs$ MRI & Field modulation can suppress optical background \tnote{\emph{i}}\\
\hline

\end{tabular}

\begin{tablenotes}
\begin{footnotesize}
\item[a]{Power reduction defined as the ratio between the power required by single primary mode and dual-mode approach to achieve certain SNR.} 
\item[b]{Extrapolated from experimental number to $d = $ 3mm, using loss coefficient 13.5 cm$^{-1}$ for 532nm laser and 12.1 cm$^{-1}$ for 650nm fluorescence in a scattering media  (See \cite{SOM}Section 1A, diamond mass $\sim$ 40mg). The coefficients are calculated based on fatty tissue data in \cite{jacques2013optical}.} 
\item[c]{Extrapolated from \cite{SOM} Fig. 2, and taking a depth of 15mm.}
\item[d]{Taken from the dual-mode experiment demonstrated in \zfr{dual_image} (See \cite{SOM}Section 1A).}
\item[e]{Extrapolated from the experiment demonstrated in \zfr{dual_image}, assuming $\mathrm{FOV} = $ (6.4mm)$^2$ (Details see \cite{SOM}Section I.B.).}
\item[f]{Obtained from \zfr{fast_image} when assuming $(1-s) = 0.9\times10^{-3}$.}
\item[g]{The SNR gain in the third mode is defined as the ratio between dual-mode SNR and optical SNR.}
\item[h]{Based on gyromagnetic ratio of $\Cs$, $G_{max}$ can be up to 60 T/m in \cite{fischer2000segment}, and $\tau \sim T_2 \approx$ 1ms. }
\item[i]{See \zfr{bg_spp}}
\end{footnotesize}
\end{tablenotes}

\end{threeparttable}

\caption{{Dual-mode imaging regimes} \normalfont{from} a combination of optics and MRI. Regime I and II consider optics and MRI being primary imaging modes in scattering and scattering-free media respectively. Regime III considers sampling in both imaging dimensions as elucidated in the algorithm of dual-mode imaging. Red color indicates parameters where image enhancements can be gained via dual-mode combination. For clarity, the variables here refer to $\gamma_0$, ratio between optical SNR and MR SNR at $d = 0$;  $P_i$ incident power density; $\Delta x$ pixel size; $\alpha_i$, $\alpha_\epsilon$ incident and emission light loss coefficient, including scattering and attenuation; $d$ object depth; $T$ imaging time; $\Delta x_{\mathrm{optics}}$, $\Delta x_{\mathrm{mri}}$, pixel size for optical and MRI; FOV, field of view; $s$, sparsity, the proportion of dark pixels.}
\label{3regime_table1}
\end{table*}

\T{\I{Conclusions}} -- We have demonstrated a method for dual-mode optical and MR imaging in diamond microparticles. Our approach relied on optically fluorescent \NV centers that simultaneously spin-polarize $\Cs$ nuclei, making the particles "bright'' under MR imaging. We discussed means by which the two modalities can be combined exploiting complementary advantages for scattering-free and high resolution imaging. We finally proposed methods for image acceleration exploiting the Fourier conjugacy inherent to optics and MRI.

\T{\I{Acknowledgments}} -- We gratefully acknowledge support from NSF under GOALI 1903839 (JAR, CAM), ONR under N00014-20-1-2806 (AA), NIH under NCI No. R43CA232901 (OS) and P41EB013598 (DBV). In addition CAM acknowledges support from Research Corporation for Science Advancement through a FRED Award; andNSF CREST IDEALS under NSF-HRD-1547830. J.H.W. acknowledges NIH 1S10RR013871-01A1 for funding the 400MHz MRI instrumentation.

%\bibliography{Biblio}

\bibliographystyle{apsrev4-1}

\input{hyperMRI-5.bbl}
\pagebreak
\clearpage
\onecolumngrid
%\begin{widetext}
\begin{center}
\textbf{\large{\textit{Supplementary Information} \\\smallskip
\bluetitle{Background-free dual-mode optical and $^{13}$C magnetic resonance imaging in diamond particles}}}\\
\hfill \break
\smallskip
X. Lv,$^{1}$ J. H. Walton,$^{2}$ E. Druga,$^{1}$  F. Wang,$^{1}$ A. Aguilar,$^{1}$ T. McKnelly,$^{1}$ R. Nazaryan,$^{1}$ F. L. Liu,$^{3}$ L. Wu,$^{1}$ O. Shenderova,$^{4}$\\ D. B. Vigneron,$^{5}$ C. A. Meriles,$^{6}$ J. A. Reimer,$^{7}$ A. Pines$^{1}$, and A. Ajoy$^{1,\BRd{\ast}}$\\
\smallskip
\emph{${}^{1}$ {\small Department of Chemistry, and Materials Science Division Lawrence Berkeley National Laboratory University of California, Berkeley, California 94720, USA.}}
\emph{${}^{2}$ {\small Nuclear Magnetic Resonance Facility, University of California Davis, Davis, California 95616, USA.}}
\emph{${}^{3}$ {\small Department of Electrical Engineering and Computer Sciences, University of California, Berkeley, California 94720, USA.}}
\emph{${}^{4}$ {\small Adámas Nanotechnologies, Inc., 8100 Brownleigh Dr, Suite 120, Raleigh, NC, 27617 USA.}}
\emph{${}^{5}$ {\small Department of Radiology and Biomedical Imaging, University of California—San Francisco, San Francisco, California 94158, USA.}}
\emph{${}^{6}$ {\small Department of Physics and CUNY-Graduate Center, CUNY-City College of New York, New York, NY 10031, USA.}}
\emph{${}^{7}$ {\small Department of Chemical and Biomolecular Engineering, and Materials Science Division Lawrence Berkeley National Laboratory University of California, Berkeley, California 94720, USA.}}
\end{center}

\twocolumngrid

\beginsupplement
%{ \hypersetup{linkcolor=darkred}
%\tableofcontents}

\tableofcontents

\section{Materials}

\section{Three imaging regimes}
The combination of optical and MR imaging, acquired simultaneously from diamond particles consisting of NV centers and hyperpolarized $^{13}$C nuclei, allows for enhancements in imaging in several different contexts. In particular, we considered three regimes of imaging (Table S1) :

\begin{table*}[!ht]
\label{3regime_table}
\begin{threeparttable}

\begin{tabular}{ |L{3.9cm}||L{4.2cm}|L{4cm}|L{4cm}|   }

 \hline
 %\multicolumn{4}{|c|}{Three imaging regimes} \\
 %\hline
Particular specifics	& \multicolumn{1}{|c|}{Regime \RomanNumeralCaps{1} }& \multicolumn{1}{|c|}{Regime \RomanNumeralCaps{2}} & \multicolumn{1}{|c|}{Regime \RomanNumeralCaps{3}}\\
	\hline
   Primary mode  & Optical imaging & Hyperpolarized MRI &  Accelerated dual-mode  \\
   \cline{1-3}
   Secondary mode & Hyperpolarized MRI & Optical imaging & imaging protocol \\
 \hline
Operating regime   &  \cellcolor{lR} Deep in scattering media & \cellcolor{lC}Shallow or in non-scattering media & \cellcolor{lC}Shallow or in non-scattering media\\
 \hline
Primary mode SNR &  \cellcolor{lC} $\propto \gamma_0 \cdot e^{-\alpha_i d}e^{-\alpha_\epsilon d}$ & \cellcolor{lC} $\propto {P_{i}} (\Delta x_{\mathrm{mri}})^2\cdot \sqrt{T}$ &  \cellcolor{lR} \\

Dual mode SNR & \cellcolor{lC} $ \propto e^{-\alpha_i d}$ & \cellcolor{lR} $ \propto \frac{\mathrm{FOV}}{\Delta x_{\mathrm{optics}}^2}  {P_{i}} (\Delta x_{\mathrm{optics}})^2\cdot \sqrt{T/\frac{\mathrm{FOV}}{\Delta x_{\mathrm{optics}}^2}}$ &  \multirow{-2}{*}{$ \propto \sqrt{T/(1-s)^{1/2}}$ \footnotemark[3] \cellcolor{lR} }\\

SNR gain & $1/\gamma_0 \cdot e^{\alpha_\epsilon d}$ & $\frac{\sqrt{FOV}\Delta x_{\mathrm{optics}}}{\Delta x_{\mathrm{mri}}^2}$ &    $\propto 1/(1-s)^{1/4} $ \footnotemark[7] \\
Primary mode SNR (example)  & \cellcolor{lC}4.4$\times10^{-3}\sqrt{\mathrm{Hz}}  $  \footnotemark[2] & \cellcolor{lC} 0.63$\sqrt{\mathrm{Hz}}$ \footnotemark[4] &  \cellcolor{lR} \\ 
Dual mode SNR (example) & \cellcolor{lC}1.1$\times10^{-2}\sqrt{\mathrm{Hz}}  $ \footnotemark[2] &  \cellcolor{lC} 6.3$\sqrt{\mathrm{Hz}}$ \footnotemark[5] &  \multirow{-2}{*}{\cellcolor{lR}47.5$\mathrm{\sqrt{Hz}}$ \footnotemark[4] }\\
SNR gain (example)& \cellcolor{lC}2.5 & \cellcolor{lR} 10 & \cellcolor{lC}5 \footnotemark[6] \footnotemark[7] \\
 \hline
Primary mode resolution $\delta x_p$ & $ \propto d$ & $\delta x_{\mathrm{mri}}$ & \\
Dual mode resolution $\delta x_d$  & $\sim \frac{1}{\gamma G_{max} \tau}$ & $\delta x_{\mathrm{optics}}$ &\multirow{-2}{*}{ $\delta x_{\mathrm{optics}}$}\\
Resolution gain $\frac{\delta x_p}{\delta x_d}$ & $ \propto d$  & $\frac{\delta x_{\mathrm{mri}}}{\delta x_{\mathrm{optics}}}$ & N.A.\\
Resolution limit &  $\sim \frac{1}{\gamma G_{max} \tau}$ ($\sim$ 1$\mu$m \footnotemark[8])& $ \frac{\lambda}{2} $ & $ \frac{\lambda}{2} $\\
Primary mode resolution (example) & \cellcolor{lC}1200$ \mu$m \footnotemark[3] & \cellcolor{lC} 640$\mu$m \footnotemark[4] & \cellcolor{lR}  \\ 
Dual mode resolution (example) & \cellcolor{lC} 640$\mu$m \footnotemark[4] &  \cellcolor{lR} 40$\mu$m \footnotemark[4] & \multirow{-2}{*}{\cellcolor{lR} 40$\mu$m \footnotemark[4]}\\
Resolution gain (example)& \cellcolor{lC} $ 1.875$ &  \cellcolor{lR} 16 &  N.A.\\
 \hline
Power reduction \footnotemark[1] & $1/\gamma_0 \cdot e^{\alpha_\epsilon d}$ & $\frac{\sqrt{\mathrm{FOV}}\Delta x_{\mathrm{optics}}}{\Delta x_{\mathrm{mri}}^2}$ & $\propto 1/(1-s)^{1/2}$\\
Power reduction (example) & \cellcolor{lC} 2.5 & \cellcolor{lC} 10 &  \cellcolor{lR} 25 \footnotemark[6] \\
\hline
Background suppression & MRI can suppress optical background & Minimal background in $\Cs$ MRI & Field modulation can suppress optical background \footnotemark[9]\\
\hline

\end{tabular}

\begin{tablenotes}
%\footnotetext[1]{Example case refers example case of $d =0$ for shallow regime and $d =5$mm for deep regime.} 
%\footnotetext[2]{This is the SNR gain with respect to the single mode imaging, namely the gain of primary mode versus the secondary mode.} 
%\footnotetext[3]{This is resolution that is derived based on experiments in Figure 1 and extrapolated to the example case of $d =5$mm for deep regime.}

%\footnotetext[1]{Power reduction is calculated based on power to achieve same SNR with primary mode only or with dual-mode imaging.} 
%\footnotetext[2]{Extrapolated from experimental number (See Section 1A, diamond mass $\sim$ 40mg) to $d = $ 3mm, using loss coefficient 13.5 cm$^{-1}$ for 532nm laser and 12.1 cm$^{-1}$ for 650nm fluorescence in a scattering media. The numbers are calculated based on fatty tissue data in \cite{jacques2013optical}.} 
%\footnotetext[3]{Extrapolated from \zfr{resolution}, and takes a depth of 15mm.}
%\footnotetext[4]{Taken from the dual-mode experiment demonstrated in Figure 1(See Section 1A, diamond mass $\sim$ 40mg).}
%\footnotetext[5]{Extrapolated from the experiment demonstrated in Figure 1, assuming $\mathrm{FOV} = $ (6.4mm)$^2$ (Details see Section I.B.).}
%\footnotetext[6]{Obtained from Figure 4 when $(1-s) = 0.9\times10^{-3}$.}
%\footnotetext[7]{The SNR gain is defined as the ratio between dual-mode SNR and optical SNR.}
%\footnotetext[8]{Based on gyromagnetic ratio of $\Cs$, $G_{max}$ can be up to 60 T/m in \cite{fischer2000segment}, and $\tau \sim T_2 \approx$ 1ms. }
\begin{footnotesize}
\item[a]{Power reduction defined as the ratio between the power required by single primary mode and dual-mode approach to achieve certain SNR.} 
\item[b]{Extrapolated from experimental number to $d = $ 3mm, using loss coefficient 13.5 cm$^{-1}$ for 532nm laser and 12.1 cm$^{-1}$ for 650nm fluorescence in a scattering media  (See Section \ref{regime1}, diamond mass $\sim$ 40mg). The coefficients are calculated based on fatty tissue data in \cite{jacques2013optical}.} 
\item[c]{Extrapolated from mian Figure 1, and taking a depth of 15mm.}
\item[d]{Taken from the dual-mode experiment demonstrated in main Figure1 (See Section \ref{regime1}).}
\item[e]{Extrapolated from the experiment demonstrated in main Figure 1, assuming $\mathrm{FOV} = $ (6.4mm)$^2$ (Details see Section \ref{regime2}).}
\item[f]{Obtained from main Figure 4 when assuming $(1-s) = 0.9\times10^{-3}$.}
\item[g]{The SNR gain in the third mode is defined as the ratio between dual-mode SNR and optical SNR.}
\item[h]{Based on gyromagnetic ratio of $\Cs$, $G_{max}$ can be up to 60 T/m in \cite{fischer2000segment}, and $\tau \sim T_2 \approx$ 1ms. }
\item[i]{See main Figure 3.}
\end{footnotesize}
\end{tablenotes}

\caption{
%\textbf{Three imaging regimes} are presented and compared in the table. In the table, the red color highlights the best performed regime among the three, whereas blue marks the regular ones. The cells that are not marked with colors imply no direct comparison. \textbf{Regime \RomanNumeralCaps{1}} exploits the fact that optics is susceptible to scattering while MRI is not. The SNR and resolution improvement is illustrated when imaging through scattering media. Note that we assume that a natural $\Cs$ MR image has a negligible background, which allows the secondary mode in this regime to suppress optical background possibly from scattering speckles or autofluorescence in biological settings. In \textbf{Regime \RomanNumeralCaps{2}}, an approach of rastering the laser beam to selectively polarize diamonds in single pixels down to optical resolution provides both resolution and SNR gain. \textbf{Regime \RomanNumeralCaps{3}} explores the Fourier reciprocity to combine optical and MR imaging modalities, and provides a high SNR, high resolution, fast imaging approach. Variables in the table: $\gamma_0$, ratio between optical SNR and MR SNR at $d = 0$;  $P_i$ incident power density; $\Delta x$ pixel size; $\alpha_i$, $\alpha_\epsilon$ incident and emission light loss coefficient, including scattering and attenuation; $d$ object depth; $T$ imaging time; $\Delta x_{\mathrm{optics}}$, $\Delta x_{\mathrm{mri}}$, pixel size for optical and MRI when performed as a single mode imaging; FOV, field of view; $s$, sparsity, the portion of dark pixels.
\textbf{Dual-mode imaging regimes} from a combination of optics and MRI. Regime I and II consider optics and MRI being primary imaging modes in scattering and scattering-free media respectively. Regime III considers sampling in both imaging dimensions as elucidated in the algorithm of dual-mode imaging. Red color indicates parameters where image enhancements can be gained via dual-mode combination. For clarity, the variables here refer to $\gamma_0$, ratio between optical SNR and MR SNR at $d = 0$;  $P_i$ incident power density; $\Delta x$ pixel size; $\alpha_i$, $\alpha_\epsilon$ incident and emission light loss coefficient, including scattering and attenuation; $d$ object depth; $T$ imaging time; $\Delta x_{\mathrm{optics}}$, $\Delta x_{\mathrm{mri}}$, pixel size for optical and MRI; FOV, field of view; $s$, sparsity, the proportion of dark pixels.}
\end{threeparttable}

\end{table*}

\subsection{Regime I: Optical imaging as the primary imaging mode}	
\label{regime1}
    
Let us consider the first regime where optical imaging is the primary mode employed the diamond particles, but one has the additional capability to acquire the image through MRI. 

\textbf{Operating Regime -- } 
This mode corresponds to the operating regime where optical imaging is the primary mode, and when optical scattering prevails, an additional MR modality provides better SNR, resolution, imaging depth and imaging power reduction. In non-scattering cases, optics preserves advantages over MRI if objects are shallow. However, in scattering media, the SNR and resolution of the optics deteriorates rapidly with increasing of depth (depth $> \sim$ mm).  On the other hand, MRI via RF signal acquisition that is independent of depth, offers a complementary approach to improve image qualities and to extend the accessible depth of this imaging regime.

Under the assumption of objects buried in scattering media, the \textit{critical depth} where MRI shows equal efficacy with optics depends on the scattering properties. For instance, if we take the scattering media describe in Table I, the critical depths are $\sim$ 2 mm for SNR and $\sim$ 6 mm for resolution (details see Section \ref{snr_section} and \ref{res_section}). This suggests that MR starts to play a critical role when imaging objects are below such critical depths.

%objects are shallow (depth $< \sim$ mm). In this regime, optical imaging often presents advantages over MRI in SNR as well as resolution when same amount of optical power is applied. We analyze the two imaging modes in the following perspectives in order to draw a comparison.

\textbf{Signal to noise -- }
We here analyze the SNR enhanced by the secondary mode -- MRI, which is acquired by RF wave that immunes to scattering by media, such as tissue. %Fast decay of SNR happens when light is scattered by media, which constitutes the major challenge for optics. 
The SNR of an image is determined by factors associated with the specific experimental settings. %The expressions of such dependency allow us to normalize the SNR for two modalities so as to fairly compare them and determine the primary mode in this regime.
We first study the expressions of the optical SNR in terms of incident power density $P_i$, pixel size $\Delta$x, sample depth $d$, as well as average time $T$. Within the linear regime of power dependency, the image SNR of optics is proportional to: 
        \begin{equation}
            SNR \propto \gamma_0 P_i (\Delta x)^2\cdot e^{-\alpha_i d}e^{-\alpha_\epsilon d}\sqrt{T}
            \label{opt_snr}
        \end{equation}
where $\gamma_0$ is the SNR ratio between optics and MRI at 0 depth, and $(\Delta x)^2$ accounts for the cumulative signal in a pixel area, and $e^{-\alpha_i d}$, $e^{-\alpha_\epsilon d}$ are factors representing ``round-trip" scattering loss of incident and emission light. In the shot noise limited regime caused by attenuation and scattering, SNR can be increased by averaging, leading to the factor $\sqrt{T}$ in the expression. Assuming the luminescence is applied uniformly across the entire sample, we can use the total optical power $P_{tot}$ and the imaging field of view FOV to substitute $P_i$ and rewrite SNR as: $SNR \propto \gamma_0 \frac{P_{tot}}{FOV} (\Delta x)^2\cdot e^{-\alpha_i d}e^{-\alpha_\epsilon d} \sqrt{T}$. In the experiment of main Figure 1A, the SNR of the optical imaging in Figure 1E is 12, as $P_i$=320 mW/mm$^2$ , $\Delta x$=40 $\mu$m, $d$=0, $T$ = 0.1s.
    
    Similarly, for an optically hyperpolarized MR image, we find the SNR expression is similar to an optical image in terms of nuclear polarization to be $P_{nuclei}$: $ SNR \propto {P_{nuclei}} (\Delta x)^2\cdot e^{-\alpha_i d}\sqrt{T} $
        Typically, in the linear regime of optical power dependence:
        \begin{equation}
            SNR \propto {P_{i}} (\Delta x)^2\cdot e^{-\alpha_i d}\sqrt{T}
        \label{mri_snr}
        \end{equation}
%        And if we normalize it with respect to the acquisition time T, we have:
%        \begin{equation}
%            SNR/\sqrt{T} \propto {P_{i}} (\Delta x)^2\cdot e^{-\alpha_i d}
%        \label{nsnr}
%        \end{equation}
where the $e^{-\alpha_i d}$ factor represents the scattering loss of the incident light. In the experiment of Figure 1, $ P_{nuclei}\sim 0.3\%$, $P_i$=80 mW/mm$^2$  ($P_{tot}$=1W), $\Delta x$= 160$\mu$m, $d$=0, $T$ = 0.8s. The image in Figure 1F presents a SNR of 4.
    
    In order to fairly compare the SNR of the two modes, we need to rescale it based on power and normalize it with respect to imaging time. For example, under the same excitation power with MR imaging (80 mW/mm$^2$), the optical SNR in main Figure 1E is rescaled to 3.0. This corresponds to an optical SNR of 9.5$\sqrt{\mathrm{Hz}}$ when further normalized by acquisition time (0.1 s). Similarly, normalizing the SNR of MRI, we obtain 0.63$\sqrt{\mathrm{Hz}}$, with consideration of 40s dead time to hyperpolarize the sample. In this case, the ratio between the two SNRs is $\gamma_0 =$ 15.
    
    Now we analyze how MRI can provide a SNR gain when $d >  \sim$mm. Comparing Equation \ref{opt_snr} with Equation \ref{mri_snr}, we see a SNR gain of $1/\gamma_0 \cdot e^{\alpha_\epsilon d}$ when implementing the secondary mode. With $d = 0$ as a benchmark (in Figure 1), we derive the SNR of the scenario where $d \neq 0$ based on the SNR formulas above in Equation \ref{opt_snr}.
    %For optical imaging, when objects are buried at depth of $d$, the SNR will be deteriorated by a factor of $e^{-\alpha_i d}$ due to incident light scattering and attenuation, as well as a factor of $e^{-\alpha_\epsilon d}$, which accounts for the ``round-trip" loss. However for MRI, SNR is only decreased by a factor of $e^{-\alpha_i d}$, which corresponds to a factor of $e^{-\alpha_\epsilon d}$ less than optics. (Detailed analysis can be found in supplementary information Section V.A. and V.C.). This implies that although both modes suffer from scattering and attenuation of the illumination light, MR modality is the optimal choice in this regime.
If we take objects buried under 3 mm as an example, assuming we have the same experimental setup, the ratio of time-normalized SNR between MRI and optics  is $1/\gamma_0 \cdot e^{\alpha_\epsilon d} = 2.5$. %Here $\gamma_0 =$15 represents the ratio of normalized SNR between optics and MRI.

    %As a result, in the shallow objects regime, optics as the primary imaging mode normally presents 1 order of magnitude higher SNR than MRI according to the calculation above. 
    
\textbf{Resolution -- }
    In this regime, the MR modality compensates the resolution degradation of the primary mode optics. %This attributes to the fact that optical and MRI resolutions are determined in different ways, mainly due to their fundamental discrepancy in imaging principle. 
The resolution of MRI depends on gyromagnetic ratio $\gamma$, maximum gradient of the imager $G_{max}$, and echo time $\tau$: , 
        \begin{equation}
\delta x \propto \frac{1}{\gamma G_{max} \tau}
\label{mri_res}
        \end{equation}
which is irrelevant to object depth. However, optical resolution strongly depends on scattering distortion; $\delta x$ can be simulated by integrating the Wigner function assuming the scattering media is homogeneous, and the linear approximation holds in this regime (see \zfr{resolution}):
        \begin{equation}
            \delta x \propto d
            \label{opt_res}
        \end{equation}
        
%Notably, $\delta x$ for MRI is not restricted by object depth $d$, mainly due to the fact that the RF signal penetrates better than optical wavelength signal.

Similarly, taking the ratio between Equation (\ref{mri_res}) and Equation (\ref{opt_res}), the resolution gain enabled by MRI scales linearly with $d$. In an example case of $d$ = 15mm in a typical media (see Section IIIC), MRI produces better resolution (640$\mu$m) than optics (1200$\mu$m). 

    The ultimate resolution limit of the two modalities are different. Optical imaging is set by the diffraction limit $\delta x = \frac{\lambda}{2} \sim  0.3 \mu$m, where $\lambda$ is the wavelength of the fluorescent light. In the practical case (Figure 1E), optical pixel size $\Delta x=40 \mu$m, which is far from the diffraction limit. This is due to low N.A. of the lens and CCD resolution limit. In contrast, as a Fourier imaging technique, MRI resolution is not limited by electromagnetic wave diffraction, though in practical settings, it is normally restricted to low field gradients. Resolution of 1$\mu$m is achievable for $\Cs$ imaging if the gradient is strong enough as in \cite{fischer2000segment}. Note that here we only consider imaging solids where diffusion is not operative.%For the MR image in Figure 1E, $\delta x=640 \mu$m, given $G_{max}$ = 950 mT/m. Note that the pixel size $\Delta x= 160 \mu$m was a result of zero-filling and smoothening from physical resolution $\delta x$.

    %Consequently, in this regime, better SNR as well as higher resolution bequeaths optics advantages over MRI.
    
    \textbf{Imaging Power --}
We anticipate imaging power reduction by dual-mode imaging in comparison with single mode optics. Imaging power is normally associated with desired SNR. If one expects a certain SNR of the final image, the corresponding imaging power required can be derived based on Equation (\ref{opt_snr}) and Equation (\ref{mri_snr}). The imaging power reduction is $1/\gamma_0 \cdot e^{\alpha_\epsilon d}$. 

Without changing the experimental setting as in main Figure 1, to characterize the incident optical power of optics and MRI on an equal footing, we normalize them to the same SNR of 1$\sqrt{\mathrm{Hz}}$. The imaging power required for optics and MRI to achieve such SNR is $P_i$=8.4 mW/mm$^2$ and 127 mW/mm$^2$ respectively when $d =$0. At depth of  $d = 3$mm, the ratio between imaging power required for optics and MRI to achieve SNR of 1 is 2.5. This entails that MR modality can provide power reduction in scattering media.
    
\textbf{Background Suppression --}
In this regime, the optical background can be suppressed by MRI considering a minimal background of $\Cs$ imaging. In usual cases, background suppression enables us to remove the time-invariant background by modulating the actual signal. As demonstrated in main Figure 3, suppression of the optical background by a factor of 5, and MRI by a factor of 2 is achieved. However, when scattering predominants, the optical modulation may degrade significantly while the secondary mode MRI remains effective. In practice, the $\Cs$ background is negligible due to low natural abundance of $\Cs$ and low gyromagnetic ratio, so implementing dual-mode imaging can generate background free MRI in this regime. Note that optical background suppression by signal modulation is still valid when scattering is insignificant. It can be applied in noisy settings especially when biological samples emit autofluorescence. In principle, 100 folds of background can be eliminated via our method\cite{sarkar2014wide}.

\subsection{Regime II: MRI as the primary imaging mode}
\label{regime2}
Now we consider MR imaging as the primary mode, with additional ability to acquire the image through optical means. %In this case, scattering largely deteriorates optical SNR and resolution, while MRI demonstrates advantages when same amount of power is applied.
In this regime, we can employ a pixel-by-pixel excitation to selectively hyperpolarize diamonds within a specific pixel. Imaging such hyperpolarized samples via MR allows us to boost the SNR and resolution of MR images.

\textbf{Operating Regime -- } 
    %We first note that the gains from MRI arise in the operating regime where MR is required as a primary imaging mode, but augmented with optical resolution by focusing down the excitation beam. 
    In this regime, MR measures signals from optically resolved pixels at each time and recomposes the signal to form a complete image. To ensure effective high resolution optical illumination, objects have to be shallow (depth $< \sim$ mm) or in non-scattering media.
    
%    objects are buried deep under media (depth $> \sim mm$).
%    
% The critical depth where MRI shows equal efficacy with optics is $\sim$ 0.4 mm for SNR (\zfr{snr_ratio}) and $\sim$ 6 mm for resolution (\zfr{resolution}). MR starts to play the primary role when imaging objects are below such critical depth.
%    
%    We compare the optical and MR imaging modes in the following aspects in order to show that MR is a suitable imaging approach in this regime: 
    
\textbf{Signal to noise -- }
The SNR of MR modality in this regime can be enhanced by optics via focusing the illumination beam. If we assume the sample can take certain amount of total power $P_{total}$, focusing incident light of the same power down to a smaller size $\Delta x^2$ equivalently enhances power density $P_i$ in Equation (\ref{mri_snr}) by a factor of $\frac{FOV}{\Delta x_{\mathrm{optics}}^2}$. Here $\Delta x_{\mathrm{optics}}$ is optical resolution or beam size. In the linear regime of optical power dependence, the SNR of a MR image is improved by the same factor with power density enhancement. In addition, we should also take into account that pixel size $(\Delta x)^2$ is reduced by a factor of $\frac{\Delta x^2_{\mathrm{optics}}}{\Delta x^2_{\mathrm{mri}}}$, resulting in less signal to be collected in one measurement. Here $\Delta x_{\mathrm{mri}}$ represents the resolution of a MR image acquired in single mode. Last, scanning through the entire FOV with a fine resolution takes longer time than wide field imaging. The acquisition time $\sqrt{T}$ extends by a factor of $\frac{FOV}{\Delta x_{\mathrm{optics}}^2}$, as more measurements need to be made. Combining all three factors, the SNR improvement in this regime is $\frac{\sqrt{FOV}\Delta x_{\mathrm{optics}}}{\Delta x_{\mathrm{mri}}^2}$.

 If we take the experiment demonstrated in main Figure 1 as an example where $FOV = (6.4$mm$)^2$, and optical $\Delta x_{\mathrm{optics}}=$ 40$\mu$m, MR $\Delta x_{\mathrm{mri}}=$ 160$\mu$m, this imaging regime  empowers up to 10 times SNR improvement. Note that in our assumption, $P_{total}$ stays the same so that $P_i$ is increased, and so as SNR. This holds if the sample takes the incident power as a whole or the heat conductivity is high so that the optical power distributes to the entire sample. In practical cases, we may not be able to increase the power density $P_i$ by that much due to a local power threshold, and the actual SNR improvement may be lower than $\frac{\sqrt{FOV}\Delta x_{\mathrm{optics}}}{\Delta x_{\mathrm{mri}}^2}$.

\textbf{Resolution --}  %Optical image resolution also deteriorates rapidly with increasing thickness of the scattering media. We can approximately write $\delta x \propto d$, and the slope is $\sim$ 0.8 mm/mm (\zfr{resolution}). Relative to optics, MR image resolution will remain the same with the case where $d = 0$. 
    %Similarly, taking an example case of $d$ = 5mm, MRI resolution being 640$\mu$m is comparable to 400$\mu$m resolution of optics.
In this regime, MR modality is augmented with optical resolution by focusing down the excitation beam. When optics focuses on certain pixels, only $\Cs$ nuclei within this pixel are hyperpolarized. Subsequently we acquire the MR signal only from this pixel with optical resolution. After raster scanning the beam across the entire FOV, we can reconstruct the image with the signals obtained. %In this regime, the image possesses optical resolution, which is ultimately limited by diffraction limit $\delta x = \frac{\lambda}{2} \sim  0.3 \mu$m. 
With such method, MRI resolution can potentially be enhanced by a factor of $\frac{\delta x_{\mathrm{mri}}}{\delta x_{\mathrm{optics}}}$. This is a factor of 16 based on the demonstrated experiment, but a factor of 2$\times 10^{3}$ can be achieved if we focus the beam to diffraction limit, namely $\Delta x_{\mathrm{optics}} = 0.3 \mu$m.
     
\textbf{Imaging Power -- }
Imaging power can be reduced via efficient power usage in this regime. Based on the analysis in ``SNR section", normalizing the SNR to 1$\sqrt{\mathrm{Hz}}$ for both single mode MRI and optically resolved MRI, we obtain that the imaging power $p_{total}$ can be reduced by a factor of $\frac{\sqrt{FOV}\Delta x_{\mathrm{optics}}}{\Delta x_{\mathrm{mri}}^2}$. This corresponds to a 10 times reduction in our experimental setting.
    
\textbf{Background Suppression -- }In this regime, we assume the $\Cs$ background is minimal because of low natural abundance in a practical setting and the image be immune from background.

\subsection{Regime III: ``Dual mode" combined optical and hyperpolarized MRI}
Finally, we analyze the regime where optical and MR imaging are combined, leveraging the fact that the two imaging modes sample in real- and k-space simultaneously. %In this case, a hybrid sampling strategy in excellent case basis corresponding to optical and MR imaging respectively can lead to enhancements over either of the imaging modalities considered separately.
In this case, an ``accelerated dual mode protocol" (described in main Figure 4) is used, and it produces faster acquisition by reducing the number of sample points.

\textbf{Operating regime -- }
    The gains of combining two imaging modalities arise in the operating regime where objects are shallow (depth  $<\sim$ mm) or in non-scattering media to allow relatively good optical SNR and resolution. This is the same as the second imaging regime we described above.
    
\textbf{Signal to noise -- }
The ``accelerated dual mode protocol" allows for higher normalized SNR than single mode optics resulting from the saving empowered by the protocol. In this regime, the final image is constructed by optical scanning. As a result, the final SNR is ultimately determined by the SNR of the optical imaging. However, the time to acquire an image is shorter by a factor of $\propto 1/(1-s)^{1/2}$ with the accelerated protocol, allowing for a factor of $ \propto (1-s)^{1/4}$ SNR enhancement.

For example, in the case illustrated in Figure 4C, if acquisition is accelerated 25 times, the normalized SNR is therefore enhanced by a factor of 5. With the same experimental setting that can achieve single mode optical SNR of 9.5$\sqrt{\mathrm{Hz}}$ (analyzed in Section IA), we expect a SNR of 47.5$\sqrt{\mathrm{Hz}}$ when combing optics and MRI.
    %The SNR of the optical imaging in Figure 1E is 9.5$\sqrt{\mathrm{Hz}}$. Here we have $P_i$=80 mW/mm$^2$, $\Delta x=40 \mu$m, d=0, T = 0.1s.
    %The actual SNR for a different setup can be derived in proportion according to the SNR relationship with the parameters.
    
\textbf{Acceleration -- }
    The immediate advantages that the combined optical and MRI imaging can enable is imaging acceleration with respect to simple raster scanning. The acquisition can be accelerated by a factor proportional to $[(1-s)]^{1/2}$. If we take $(1-s) =$ 0.5\% as an example, imaging acceleration is 25 (see main Figure 4C).
    
\textbf{Resolution -- }
    Optics determines the imaging resolution in this regime as the protocol ends with the optical sampling of real-space. As a result, the protocol can achieve optical resolution, the ultimate limit of which is defined by the diffraction $\delta x=  \frac{\lambda}{2} \sim 0.3 \mu$m.
    
\textbf{Imaging Power -- }
    Under this regime, optical power can be better focused on pixels where meaningful objects are located. Total optical power $P_{tot}$  will be reduced due to less measurements being performed in real-space so that less pixels are required to be optically addressed. The power reduction is proportional to $[(1-s)]^{1/2}$. If we use $(1-s)=0.5\%$ as an example, imaging power reduction is $\sim$ 25 times (see Figure 4).

%\textbf{Technical Challenges -- }
%    We note however that implementation in practical settings requires the following setup requirements. In addition to conventional optical and MR imaging setup, it is required to implement additional optical beam raster scanning components, for instance a galvanometer.

\section{Accelerated dual mode imaging protocol}
This section describes the details of the numerical evaluation of the fast imaging protocol in Figure 4 of the main paper. The target image for demonstration (main Figure 4B ``Target Image") is a 512 $\times$ 512 high-resolution artificial picture, higher than the target resolution (64 $\times$ 64) to mimic real objects. This real-space image corresponds to a 512 $\times$ 512 k-space dataset (shown for clarity in \zfr{k-space}B). When we perform the fast imaging protocol, a 16 $\times$ 16 region at the center of the k-space is selected to emulate MRI k-space subsampling. This sampled dataset is then zero filled to a form a 64 $\times$ 64 matrix (Figure 4B ``Sampled k-Space"), and subsequently Fourier transformed to reconstruct a blurry 64 $\times$ 64 real-space image (Figure 4B ``MR Image"). We set a threshold at 0.1 times the mean pixel value, and the pixels with values above the threshold are optically sampled to obtain Figure 4B (``Accelerated Optical Image").

\begin{figure}
  \centering
  \includegraphics[width=0.45\textwidth]{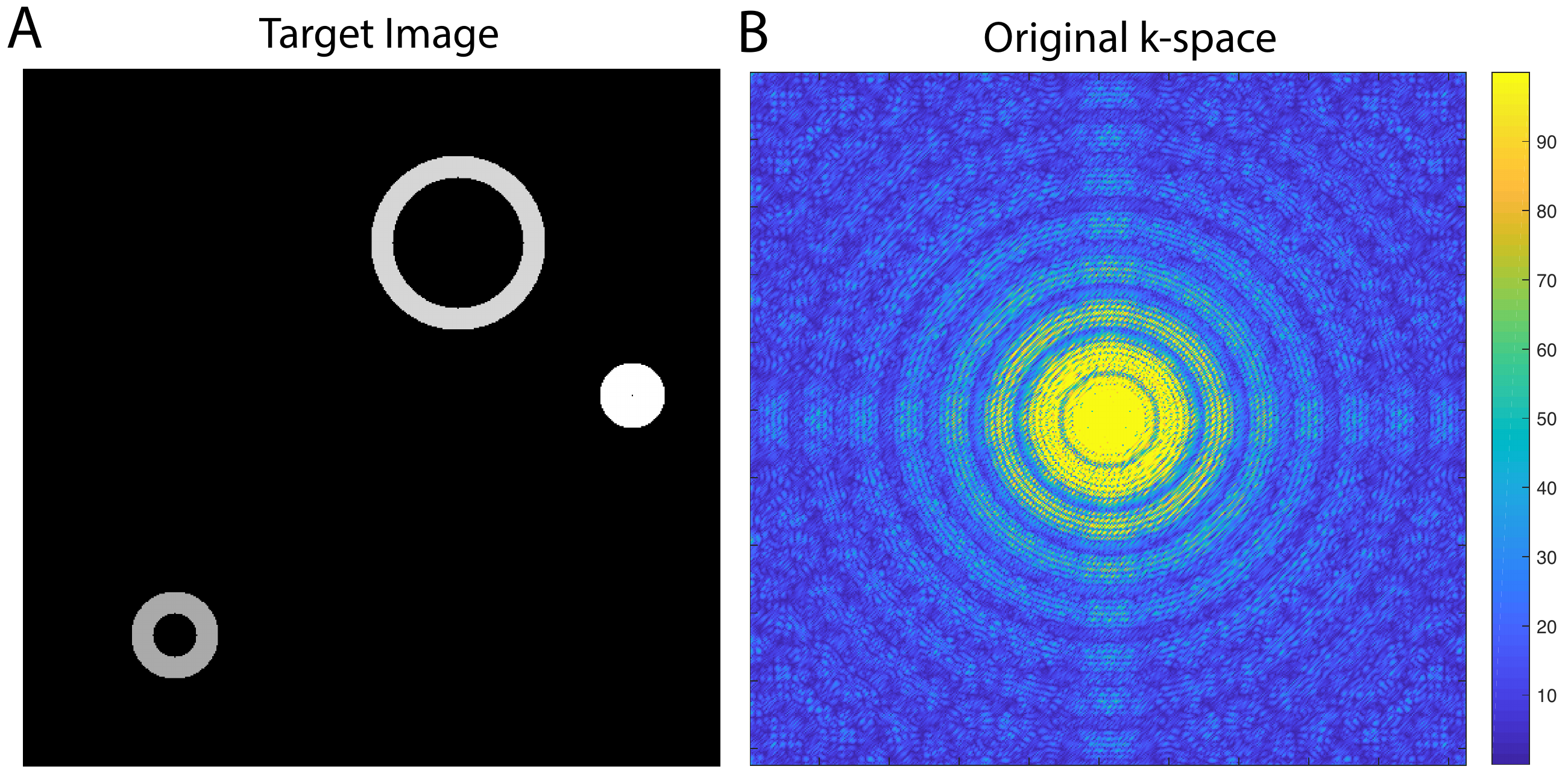}
  \caption{\textbf{Target image and corresponding k-space data} of exemplary ring-like phantom images in Figure 4 of the main paper. Both spaces have 512 $\times$ 512 pixels.}
\zfl{k-space}
\end{figure}

Using the protocol described in Figure. 4A, B, we perform numerical experiments in Figure. 4C-F with randomly generated target objects. Specifically, we randomly place ``points'' that are round in shape in a 32 $\times$ 32 FOV frame to form target images. This is relevant to diamond tracking and targeting scenarios, where diamonds are small clusters in the FOV. Each point has an intensity of 1, and the background is 0. We can control the number of``points''  inserted to the image to vary sparsity $s$. The simulated results is shown in Figure. 4C, and each data points on the plot of  correspond to an average of 30 different experiments with 30 random target images. Each individual experiment follows the protocol described above, with evaluation of acquisition acceleration and image fidelity (i.e. image correlation). We plot the optimized sampling width and total time in Figure. 4D and E.

In Figure 4F, to characterize the reconstructed image convergence towards the target as optical sampling rate increases, we evaluate the ``image correlation'' convergence curve. Image correlation is defined as $
r=\frac{\sum_{m, n} \left(A_{m n}-\overline{A}\right)\left(B_{m n}-\overline{B}\right)}{\sqrt{\left(\sum_{m, n}\left(A_{m n}-\overline{A}\right)^{2}\right)\left(\sum_{m, n}\left(B_{m n}-\overline{B}\right)^{2}\right.}}$
where $A_{m n}$ and $B_{m n}$ are the intensity of pixel $(m,n)$ in two images A and B (here the reconstructed image and target image respectively). In the simulation of optical sampling, we assume that the pixels are sampled in a certain order based on their intensity. Explicitly, from the coarse reconstructed MR image, the pixels are sorted by their amplitudes, and real-space sampling is performed with a high- to low-amplitude order. Figure. 4F indicates that sampling a few high-intensity pixels will quickly saturate the correlation curve, while smaller ones play less important role. 

\subsection{Theoretical model}

\begin{figure}
  \centering
  \includegraphics[width=0.48\textwidth]{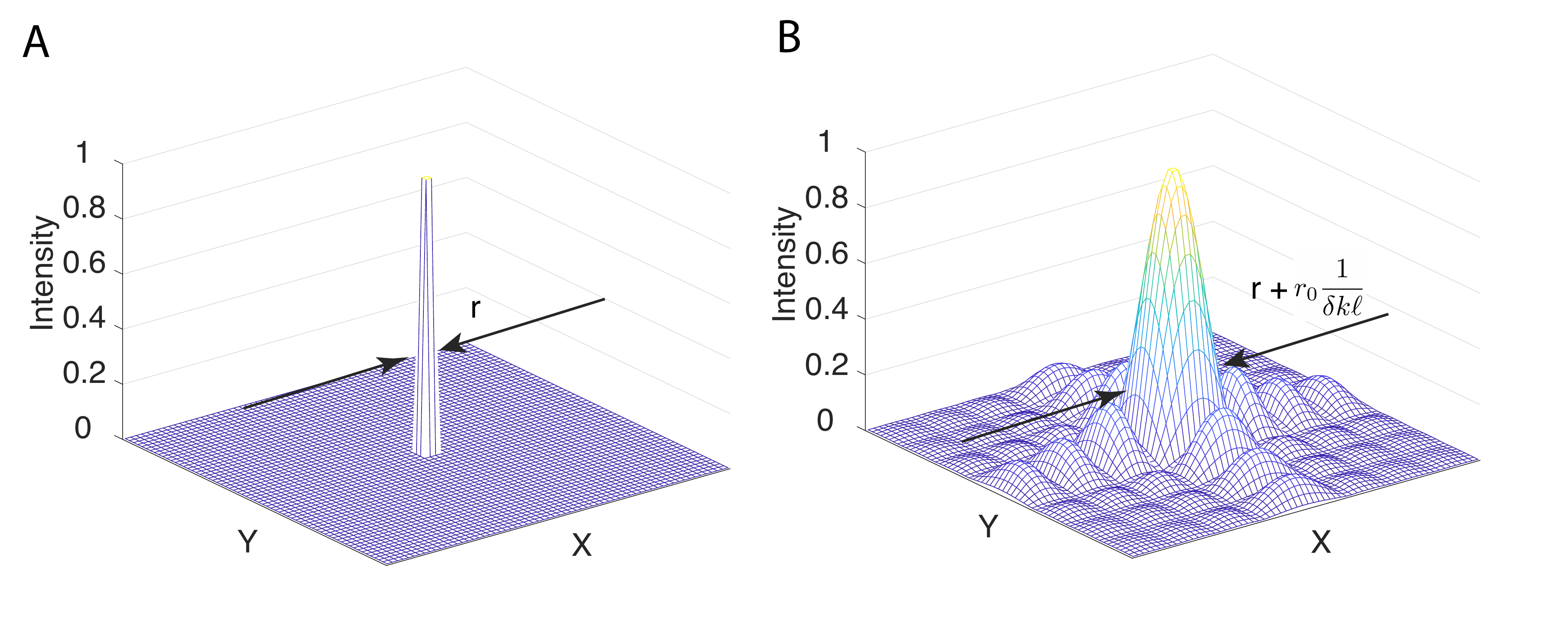}
  \caption{\textbf{Blur caused by k-space subsampling.} {(A)} The intensity map of a single point image. {(B)} The reconstructed point becomes blurred due to k-space subsampling. There are multiple peaks besides the center, following the profile of sinc functions along two dimensions ( $\sinc (W_{kx} x) \cdot  \sinc (W_{ky} y)$). Thresholding the intensity can exclude higher order peaks.}
\zfl{circle_3d}
\end{figure}

In this section, we elaborate on the scaling behavior of the algorithm and attempt to generalize the protocol to higher dimensions. 

First, let us quantify the blur caused by k-space subsampling. Assuming the original real-space image is $f(x, y)$, then the corresponding k-space function is $\hat{f}(k_x, k_y) = \mathcal{F} (f)$. Assuming a real-space pixel length of $\delta x , \delta y = $1 and a square FOV, we get $\delta k_x,  \delta k_y= \frac{1}{N}$ respectively, where $N$ is the image length in one dimension. If we consider that the sampling in k-space occurs up to $\ell$-th order, meaning that we sample a $ \ell \times \ell$ window at the center of k-space, this provides a truncated k-space image:
\begin{equation}
\hat{f}_{\ell}(k_x, k_y) = \hat{f}(k_x, k_y) \cdot \Pi \left( \frac{k_x}{W_{kx}} \right) \cdot \Pi \left( \frac{k_y}{W_{ky}} \right)
\end{equation} 
where $\Pi$ is rectangular function with length of $W_{kx} = \delta k_x \cdot \ell$ and $W_{ky} = \delta k_y \cdot \ell$. The signal transformed to real-space image then becomes:
\begin{equation}
\begin{aligned}
f_{\ell}(x,y) &= \mathcal{F}^{-1} (\hat{f}_{\ell}(k_x, k_y)) \\
&=f(x,y) * [W_{kx} \sinc (W_{kx} x) \cdot W_{ky} \sinc (W_{ky} y)]
\end{aligned}
\end{equation} 
where $*$ represents a convolution operation. For simplicity, we consider the case when there is only one single point in real-space, where $f(x,y)$ is effectively approximated by a delta function of unit pixel size in real space. General objects can be linearly constructed from this case. The effective point spread function (psf) is then the product of sinc functions $f_{\ell}(x,y)  = \sinc (W_{kx} x) \cdot  \sinc (W_{ky} y) = \sinc (x \delta k \ell) \cdot \sinc (y \delta k \ell)$. This means that truncation in k-space effectively leads to a blurring of the image in real-space (see \zfr{circle_3d}). 

In the protocol, we want to select pixels that are worth to be optically sampled by thresholding the coarse image, and this step determines the effective psf broadening. The threshold we choose here is based on the idea of excluding the shoulder of the psf (\zfr{circle_3d}B), which is approximately below 0.3 times of the maximum. With such thresholding, we will only select the center peak of the psf (in \zfr{circle_3d}B) up to a point where $x_0 = 0.75\frac{1}{\delta k \ell}$, because $\sinc (W_{kx} x_0)$ = 0.3. Thus, the effective expansion of the real space psf is $r\rt r + 0.75\frac{1}{\delta k \ell}$. The optical scan time can be estimated from the area of the psf $\pi ( r + 0.75\frac{1}{\xd k \ell})^2$, where we assume one unit area corresponds to one unit time of imaging. Considering both optical and MR imaging, total time:
\begin{equation}
T = \pi n_d \lb r + r_0\frac{1}{\delta k \ell}\rb^2 + \ell^2 
\label{T}
\end{equation} 
where $n_d$ is the number of ``points" within the FOV assuming the image is sparse enough and the points are not overlapping, and $r_0$ is conditioned on the thresholding value used in more general cases (here we use $r_0 =$ 0.75 for a threshold of 0.3 in our simulations).  For a fair comparison, we normalize the time cost against conventional imaging to get $\tau = \frac{T}{N^2}$. 

We analyze the relationship between time cost of our protocol $\tau$ with respect to the k-space sample order $\ell$ to identify the optimal sampling strategy. Note here we use $s$ to denote the image sparsity -- the portion of the zero-valued pixels. We define the ``image occupancy" to be $(1-s) = \frac{1}{N^2} \pi r^2 n_d$. Combining the equations, we have:
%\begin{equation}
%\tau = (1-s)+\frac{2r_0}{N}\frac{1}{\ell} \left( \frac{1-s}{r} \right)+\frac{r_0^2}{\ell^2}\left( \frac{1-s}{r^2}\right)+\frac{\ell^2}{N^2}
%\end{equation} 

\begin{equation}
\tau = (1-s)+\frac{2r_0}{r}\frac{1}{\ell} \left[ (1-s)N \right]+\frac{r_0^2}{r^2}\frac{1}{\ell^2}\left[ (1-s)N^2 \right]+\frac{\ell^2}{N^2}
\end{equation} 

This equation yields the fitting in main Figure. 4C. We hope to find the minimum time cost $\tau (\ell)$ analytically. Under the assumption of highly sparse images, we take approximations: $(1-s)\rt 0$. Given the definition of $1- s$, we also have $(1-s)N \rt 0$, and $(1-s)N^2\rt {\cal O}(1)$,
\begin{equation}
\tau (\ell) = A \frac{1}{\ell^2} + B\ell^2,
\end{equation}
where $A = \frac{r_0^2 (1-s)N^2}{r^2}$, and $B = \frac{1}{N^2}$. Let us now derive the corresponding optimal value for $\ell$. Taking the derivative of $\tau (\ell)$ to minimize it and identify the optimal sub-sampling size in each dimension $\ell$, we get:  $\frac{d \tau (\ell)}{d \ell} = -2A \frac{1}{\ell^3} + 2B\ell = 0$. We get $\ell_{opt} = (\frac{A}{B})^{1/4} \propto (1-s)^{1/4}$, and corresponding to an optimal time $\tau = 2(AB)^{1/2} \propto (1-s)^{1/2}$ . These formulae, in the limit of $s\rt 1$, show a good agreement with both image simulation results (data points in Figure. 4D,E) and numerical optimization results (solid line in Figure. 4D,E), that were obtained numerically without any approximations.

Let us now finally extend these calculations to the case of $d$ dimensional imaging, one can evaluate the relative imaging time under the assumption of high image sparsity:
\begin{equation}
\tau (\ell) = \frac{1}{N^d}[C(d)\pi n_d(r+r_0 \frac{N}{\ell})^d+ \ell^d] \\
\approx A_d\frac{1}{\ell^d}+B_d\ell^d
\end{equation}
where $C(d)$ is a dimension dependent constant that sets the effective $d$-dimensional volume for instance, $C(3) = \frac{4}{3}\pi$, $C(1) = \pi$, and  $A_d = \frac{r_0^d (1-s)N^d}{r^d}$, and $B_d = \frac{1}{N^d}$. We then derive optimal $\ell_{opt,d} = (\frac{A}{B})^{1/2d} \propto (1-s)^{1/2d}$, and optimal $\tau_{opt,d} = 2(AB)^{1/2} \propto (1-s)^{1/2}$. Note that the scaling of optimal relative imaging time $\tau_{opt}$ is independent of dimension $d$. 

Finally, we comment that the imaging acceleration in our case can enable a substantial and concomitant decrease of effective imaging power. This is because the protocol effectively reduces the imaging volume to be probed over. In the demonstration in Figure 4B  of the main paper, for instance, the acceleration $\tau \approx 0.073$ corresponds to a 14 times lower overall optical power on the sample.

\subsection{Comparison with compressed sensing}
\label{CompCS}
The accelerated dual-mode imaging protocol exploits the ability to simultaneously sample in real- and k-space in order to achieve acquisition acceleration and power reduction. The protocol shares some similarities with compressed sensing (CS) but differs in other aspects. As summarized in Table \ref{table1}, we compare these two methods in terms of the following perspectives:

\begin{table*}[!ht]
\begin{tabular}{ |p{4cm}||p{5cm}|p{5cm}|  }

 \hline
     & CS \cite{candes2006near, donoho2006compressed} & Dual-mode protocol\\
 \hline
Operating regime   & A known space is sparse  & Real-space is sparse \\
Principle &   RIP and $\ell_1$ minimization  & Reciprocity between k- and real-spaces \\
Technique type  & Computational & Physical \\
Subsampling space & K-space  & Both k- and real- spaces\\
Acceleration & $n/m$ (where $m=\mathcal{O}((k \log (n / k))$ or $\mathcal{O}\left((k \log (n)^{4}\right)$, $k= 1-s$ )\footnotemark[1] & $\mathcal{O}(1-s)^{1/2}$\\
\hline
Acceleration (example) &   40  \footnotemark[2] & 25 \footnotemark[3] \\
Image resolution in this paper \footnotemark[4]  & 640$\mu$m & 40$\mu$m \\
%Image resolution limit &   no fundamental limit  & 0.3$\mu$m\\
%Sparsity & required  & required \\
 \hline
\end{tabular}
\footnotetext[1]{ $m$ is the number of measurements, and $n$ is the dimension of the image vector that represents a collection of image pixels.}
\footnotetext[2]{The case corresponds to (1-s) = 0.5\%, where $s$ is sparsity in certain representation (See Section \ref{CompCS}).}
\footnotetext[3]{The case corresponds to (1-s) = 0.5\% (See main Figure 4).}
\footnotetext[4]{This is based on the experimental setting illustrated in main Figure 1.} 
\caption{\textbf{Comparison between CS and Dual-mode protocol.} In comparison to MRI CS, the ability to sample simultaneously in 2 spaces provides a new venue to acquisition speedup with additional benefit of resolution improvement. In the table, $n$, $m$, $s$ represent image dimension, number of measurements, sparsity (the portion of dark pixels) respectively. }
\label{table1}
\end{table*}

\subsubsection{Compressed sensing}
Let us first briefly review basics of CS, in order to better compare it with our protocol. CS is the origin of a major evolution in signal processing \cite{candes2006near, donoho2006compressed}. It is widely applied in photography, medical imaging, as well as astronomy, because of its ability to recover the signal from very few linear measurements.

\textbf{Operating regime --}
CS, when applied in imaging, operates in a regime where a known representation of the target image is sparse. There usually exists a sparse representation (eg. wavelet or finite-difference representation)\cite{Lustig07} for a natural image that encodes certain patterns or structures. The sparsity, which is central to CS, makes it possible to compress information in a large dimensional space into fewer measurements.

\textbf{Principle --}
The key principle that makes CS successful in MRI are random measurements satisfying restricted isometry property (RIP)\cite{donoho2006compressed} as well as $\ell_1$ minimization in a sparse representation reconstructing image space.

More specifically, let us describe the general ideas of CS quantitatively while skipping mathematical proofs. Let $\bar{u}$ be the true underlying image of interest and $\Psi \in \mathbb{C}^{m \times n}$ ($m < n$) be a sensing (sampling) matrix, where $m$ is the number ocan RIP conditionf measurements, and $n$ is the dimension of the image vector that represents a collection of image pixels. An encoding process employs a certain physical means to collect data $b \in \mathbb{C}^{m}$ that is $\Psi \bar{u}$. A decoding process is used to reconstruct the image $\bar{u}$ from $b$. 

We suppose that signal $\bar u$ is sparse in representation $\Phi$, namely $\Phi \bar{u}$ is a sparse vector. To reconstruct $\bar{u}$, one normally solves the $\ell_1$ problem when there is no noise:
\begin{equation}
\ell_{1}: \min _{u}\left\{\|\Phi u\|_{1}: \Psi u=b\right\}
\label{CS_opt}
\end{equation}

If a measurement matrix $A = \Psi \Phi^{\dagger}$ satisfies the restricted isometry property (RIP) condition, which indicates $A$ preserves the Euclidean 2-norm of S-sparse vectors, the reconstruction has a high probability to yield the true image $\bar{u}$. One of the most known cases that could make $A$ satisfy this condition is CS MRI where one collects under-sampled k-space data and minimizes $\ell_1$ norm in wavelet or finite different space. In summary, measured data $b$ with $m$ entries reconstructs $u$ which has $n$ elements.

\textbf{Acceleration --}
CS allows a factor of $n/m$ acceleration in comparison with sampling $n$ points in full. The minimum number of measurements for successful reconstruction is $m=O((1-s) \log (n / (1-s)))$ and $O\left((1-s) \log (n)^{4}\right)$, when $\Psi$ is a Gaussian random matrix and partial Fourier ensemble respectively. Practically one finds that only when $m \sim> 5(1-s)$, can the RIP condition be satisfied and a reasonably good-quality image can be reconstructed. As a result, the acceleration factor is $\sim \frac{1}{5(1-s)}$. In order to compare CS with our dual-mode accelerated protocol, we assume there is a case where real-space is considered to be the sparse space, and $(1-s)=0.5\%$. This gives us an acceleration factor of 40.

 \textbf{Image resolution --}
In the case of CS MRI, the image resolution is ultimately determined by the k-space sampling window, which is similar to a conventional MR image. The MRI spatial resolution is given by:
        \begin{equation}
            \delta x \propto \frac{1}{\gamma G_{max} \tau}
        \end{equation}
where $\gamma$ is gyro-magnetic ratio, $G_{max}$ is maximum imaging gradient, and $\tau$ is echo time. 

 \textbf{Technical challenges --}
CS in MRI is mainly a algorithmic technique that is usually easy to implement as long as one has the ability to randomly sample k-space. The major challenge however emerges when the image dimension $n$ is large, for instance 3D imaging, and large computational power is required to solve the convex optimization problem.

\subsubsection{Accelerated dual-mode protocol}

The accelerated dual-mode protocol proposed in this work takes sub-sampled k-space data by MR as a reference to guide real-space measurement by optics. This protocol warrants acquisition time saving concurrently with optical power reduction. %We consider two types of such protocols: (i) the exact protocol that is illustrated in Figure 4, relying on real-space being sparse; (ii) a more general method that performs reconstruction based on subsampled k- and real-space measurements, requiring a relaxed condition.

%Let us elaborate more on the second type protocol. In the particular case of the demonstrated dual-mode accelerated protocol (Figure 4), we take advantage of sparsity in real-space (which is likely the case for diamond when particles are loosely distributed in the image). However, sparsity of other spaces can also be exploited, such as the finite difference representation, making dual-mode imaging a more general idea to promote acceleration. 

%In practice, we can envision the protocol to sample in a sparse the finite difference representation in the following way. Subsampling in k-space enables us to reconstruct a blurry image in real-space using either inverse Fourier transform or CS. The blurry image can be used as a reference to seek high frequency ``edges" (defined as pixels where intensity changes rapidly and local gradient $\Delta I$ is above certain threshold). Then real-space sampling addresses the edges by taking the difference of adjacent pixels, leading to indirect tackling of the finite difference representation. In other words, subsampling in real-space in method (i) is now substituted by subsampling in the finite difference space. As a consequence, in this method, the requirement of sparsity in real-space is relaxed to the finite difference representation. More details of the second category protocol will be presented in the forthcoming manuscript.

\textbf{Operating regime --}
Such protocol normally presents speedups in the cases where an image is sparse in real-space. In most of the diamond particle imaging, such as tracking or targeting, the criterion is satisfied because diamonds occupy few pixels in a large FOV when we consider such applications. We note that, the preliminary image is ``generated" during the process rather than ``required" as a precondition of our protocol. Thus, while we consider optical imaging and MRI together as a complete unit, the requirement for an image that is suitable for our dual-mode protocol is only real-space sparsity.

%However, the idea of dual-mode imaging acceleration is not only limited to cases where real-space is sparse. With the type (ii) protocol, the same principle can be applied to more general scenarios where finite-difference representation of the image is sparse. The relaxed premise includes images like brain images or angiography into the operating regime of our protocol.

\textbf{Principle --}
The principle of the protocol is the reciprocity of the real- and k-spaces, where optical and MR imaging take place respectively. The fact that very few k-space points can lead to a coarse real-space model stems from the Fourier reciprocity. Such principle thus differs from CS. The divergency of the principles of CS and the dual-mode protocol may allow us to superimpose the two methods to get benefits from both. Specifically, random sampling together with convex minimization reconstruction can provide a blurry real-space image for optical real-space imaging to refine. Although the idea is straightforward, it requires careful design of the protocol and the reconstruction algorithm. More details will be in a forthcoming manuscript.

\textbf{Acceleration --}
The acquisition can be accelerated by a factor proportional to $(1-s)^{1/2}$. If we take $(1-s)=0.5\%$ as an example, imaging acceleration is 25 (see main Figure 4C). Notably, additional imaging power reduction happens concurrently with acceleration due to having less pixels necessary for optical excitation.

 \textbf{Image resolution --}
In contrast to CS, the resolution of the image reconstructed by the protocol is determined by optical resolution. As analyzed in the Section \ref{res_section}., for a mm-deep buried sample, image resolution $\delta x$ can be approximated to proportional to $d$ (\zfr{resolution}) as $\delta x \propto d$ by integrating the Wigner function assuming the scattering media is homogeneous. The ultimate resolution limit of optical imaging at $d =0$ is set by the diffraction limit $\delta x = \frac{\lambda}{2} \sim  0.3 \mu$m , where $\lambda$ is the wavelength of the fluorescent light.

 \textbf{Technical challenges --}
The method is an experimental approach compared to CS which is mainly algorithmic. The protocol relies on acquiring information in k-space first and feedforward it to real space sampling. Thus, the protocol yet requires implementation of a scanning optical imaging platform in addition to MR imaging facilities. Optical setups are normally easy to build at lower cost though.

\subsubsection{Similarities}
Apart from the differences described above, CS and dual-mode accelerated protocol share some similarities in the following ways.

 \textbf{Subsampling --}
Both methods exploit subsampling to avoid scanning across the entire imaging space and therefore gain time saving. CS MRI subsamples k-space while our protocol subsamples both Fourier conjugate spaces.

\textbf{Sparsity --}
Both methods rely on sparsity, which is the key to data compression in general. Sparsity means that when a signal is represented in a certain basis, eg. $I = \sum a_i \phi_i$, the majority of the coefficients $a_i$ are 0. This indicates that the number of linearly independent coefficients is less than the dimension of the image-space. Both CS and our protocol take advantage of the sparsity of the target image to compress the number of samples and achieve imaging acceleration.

\section{Imaging efficiency of two modalities}
In this section, we estimate and compare the imaging efficiency of the optical and MR modalities. We assume the imaging objects are randomly oriented diamond particles buried under depth of $d$ in certain media. Green incident photons pass through the media to excite the diamonds, while either an optical detector and or an MRI detector (RF coil) are utilized to image the particles. 

\subsection{Imaging efficiency of optics}
To unravel more clearly the factors that affect imaging efficiency, we consider the overall ``\I{signal path}'' from the green photon excitation to the ultimate collection of red fluorescence.

\textit{1. Green photon scattering and loss at media.} Scattering and attenuation lead to loss of photon flux. The attenuation loss can be evaluated by Beer's law:$\frac{\varphi_p}{\varphi_{m}}=\R{exp} \{ \int_{0}^{d_0}- [\alpha(d)+\mu(d)]\cdot \R{d}d\}$, where $\varphi_p$ is the photon flux of the incident pump beam, and $\varphi_{m}$ is the part that reaches diamonds. $\alpha(d)$ and $\mu(d)$ denote the attenuation and scattering coefficient of the media respectively. A typical attenuation coefficient $\alpha(d)$ for skin tissue is about 2 cm$^{-1}$ at the wavelength of 532nm\cite{Lister12}. The scattering coefficient can be written as $\mu=R\cdot \lambda^{-4}$\cite{Jacques96} for Rayleigh scattering. %, where $R=2\times 10^{-13} mm^{-3}$\cite{Jacques96}. 
We take scattering coefficient of 13.5 cm$^{-1}$ at 532nm which is calculated based on fatty tissue data in \cite{jacques2013optical}. For diamond buried under 3 mm of such media, we estimate an attenuation loss of 0.54 and scattering loss of 1.7$ \times 10^{-2}$, and hence $\eta_{o,1}=\frac{\varphi_p}{\varphi_{m}} =9.6 \times 10^{-3}$. Note that here we have ignored the losses due to the reflection at the diamond surface, which is on the order of 0.83 at near-normal angles of incidence~\cite{Clevenson15}, and can be considered negligible compare to $\eta_{o,1}$.

\textit{2. Green-to-red photon conversion.} In fact, during optical illumination, only a small portion of the incident green photons finally convert to red photons by NV centers. We now estimate this conversion rate $\eta_{o,2}$ based on Ref ~\cite{Clevenson15} . A waveguide is employed on a 0.1 ppm NV diamond crystal in this literature, however, we only take the reported single pass conversion efficiency which is relevant to our experimental scenario. We then have $\eta_{o,2}=5.5\times 10^{-5}$.

\textit{3. Total internal reflection loss.} NV centers emit in red, and a portion of the emission light can be restricted within the diamond because of total internal reflection (TIR). %The high refractive index of diamond ($n_d=2.4$) restricts red light that is within a critical angle from coming across the diamond-air interface. %Only refractive light can be possibly detected. 
The critical angle of TIR for a diamond-media interface is $\theta_{c}=\arcsin \left( \mathrm{n}_{\mathrm{m}} / \mathrm{n}_{\mathrm{d}}\right)=24.6^{\circ}$. We calculate $\eta_{o,3}$ as the portion of the light that is not confined in diamond at the first incidence. Assuming $\mathrm{n}_{\mathrm{m}}=1$, the high refractive index of diamond ($n_d=2.4$) leads $\eta_{o,3}\approx 9\times 10^{-2}$~\cite{Lesage12}.

\textit{4. Red fluorescence loss.} Optical imaging incurs ``\I{round-trip}" losses due to the need to also collect NV fluorescence. Similar to Step 1 above, one can estimate these attenuation and scattering losses. We take the attenuation coefficient at 650nm as $\sim$1 cm$^{-1}$, and the scattering coefficient as $\sim12.1 \R{cm}^{-1}$ (calculated based on fatty tissue data in \cite{jacques2013optical}). For diamonds buried 3 mm deep in such media, we calculate $\eta_{o,4}\approx 1.9 \times10^{-2}$, where attenuation loss is 0.74 and scattering loss is $ 2.6 \times10^{-2}$.

%We note that apart from signal losses, scattering can also lead to blur (and loss of resolution) in the images. An experiment \cite{hodgkinson1994point} shows that, imaging through typical blood, the full width half maximum (FWHM) of the point spread function (PSF) is $\sim$ 200 $\mu$m when passing through even a very thin layer (500$\mu$m). Assuming FWHM would roughly scale exponentially with depth, in a 2mm thick tissue, a point light source will be blurred to $\sim$cm. MR imaging, in contrast, can provide high resolution even in scattering environments.

\textit{5. Geometric effects in light collection.} The finite numerical aperture (NA) of the detection optics restricts the total number of photons that can be collected. Only fluorescence light within a solid angle $\Omega$ can be captured by the objective lens. We have:
\begin{equation}
\Omega=2\pi(1-\cos(\theta_{max})),
\end{equation}
where $\theta_{max}=\arcsin(\R{NA}/ n_d)$. Assuming that the fluorescence emission is initially spatially isotropic, the detection efficiency can be expressed as $\eta_o^5=\Omega/4\pi$. Considering $\R{NA} = $ 0.18 in our experiment, we estimate $\eta_{o,5}= $8$\times 10^{-3}$ .

\textit{6. Signal amplitude reduction by lock-in detection.} Finally, we consider the amplitude loss caused by lock-in detection in practical scenarios. Lock-in techniques are normally used when optical background presents, and the signal of interest can be discerned by the method. However, such benefit of background suppression comes at the cost of reduced overall signal amplitude. This stems from the finite modulation contrast ($<$100\%) upon application of magnetic fields (see main Figure 2A). Since only the component that can be modulated is recovered, the overall signal is weighted by the modulation contrast ($\sim$10\% typical in our experiments, as in main Figure. 2A(ii)), $\eta_{o,6}\approx 0.1$.

Putting together these factors, we estimate the approximate loss incurred from green photons leaving the light source to finally being detected as florescence. This signal-loss chain already draws a strong contrast of optical imaging with the case of MRI, since in the latter section, MRI circumvents scattering losses on the return path. In summary, for the example case of of diamond particles imaging under 3 mm, $\eta_o= {\displaystyle \prod_{i=1}^{6} \eta_{o,i}}=7.2 \times10^{-13}$.

\subsection{Imaging efficiency of MRI}
We now perform a similar analysis for the case of MR imaging. While there is indeed less loss in MRI compared to the ``round-trip" loss of optics, there are additional factors related to the inductive detection of the $\Cs$ magnetization. Once again we ignore the efficiency of the NMR detector itself (related to preamplifier gain and noise) and focus instead only on the signal chain up to the detector. %We note that in our experiments, the typical image SNR is 1.05, employing 1W laser and polarizing the spins for 40s. 

\textit{1. Green laser loss at media.} Similar to the optical imaging case, since the hyperpolarization is optically induced, we continue to incur \I{one-way} optical scattering and attenuation losses through the imaging media, corresponding to a factor $\eta_{m,1}=\eta_{o,1}=9.6 \times 10^{-3}$ as above.

\textit{2. Quantum efficiency of $\Cs$ hyperpolarization.} Let us now estimate efficiency of the $\Cs$ hyperpolarization process. In Figure 1C, the hyperpolarization signal is obtained by 1W laser irradiation with a duration of 40s; as a result, the 10mg natural abundance diamond sample is polarized to $\sim$0.3\%. The number of polarized $\Cs$ spins $n_c$ can be calculated as $n_c=\frac{m}{MW} \cdot P \cdot \epsilon \cdot N_A$, where $m$, $MW$, $P$, $\epsilon$, $N_A$ are: the mass of the diamond powder, the molecular weight of the sample, $\Cs$ polarization, $\Cs$ enrichment and Avogadro number. We obtain $n_c=5.0\times 10^{16}$. On the other hand, the total number of green photons used to pump the diamond is:
\begin{equation}
n_p=\frac{1W\times 40s}{\hbar \times \frac{3\times 10^8 \R{m/s}}{532 \R{nm}}}=1.1\times 10^{20}
\end{equation}
leading to a total efficiency $\eta_{m,2}=\frac{n_c}{n_p}=1.4\times10^{-4}$. This also indicates that approximately 7200 green photons can effectively polarize one $\Cs$ nuclear spin. We note here: our experiments suggest that when a sample is $\Cs$ enriched to 10\%, the overall polarization enhances 10 times. This result implies that the efficiency can be further boosted by at least an order of a magnitude with $\Cs$ enrichment. 

\textit{3. Geometric factors related to detection.} The sample filling factor in MR imaging plays a similar role as the geometric factors related to finite numerical aperture in optical imaging. In our experiments, the volume of the coil is approximately 2400 mm$^3$, and the typical volume occupied by the particles is 11 mm$^3$, implying a relatively low filling factor of $\eta_{f}$ = 4$\times 10^{-3}$. The MR signal scales $\propto \eta_f^{1/2}$ when the volume of the coil is fixed~\cite{hoult1976signal}. We define $\eta_{m,3} = \eta_f^{1/2}$, and we have $\eta_{m,3} =$6.3$\times 10^{-2}$. We note therefore that our filling factor is far from optimal, and can be boosted by at least an order of magnitude by using a coil that is more suitably matched to the sample of interest, or through of use of inductive coupling coil chains~\cite{hai2019wireless}.

\textit{4. Detection frequency.} We note that the hyperpolarization process delivers $\Cs$ polarization that is agnostic to detection field. However, the choice of detection field does play a role in determining the final obtained SNR, scaling as
$\propto  \left(Q \omega_{0}  \right)^{1 / 2}$ ~\cite{Hoult1978},
where $Q$ is the quality factor of the coil, and $\omega_{0}$ is the Larmor frequency under that field.

\textit{5.Detection Time.} Finally, the image SNR is also a function of the total effective $\Cs$ $T_2$ coherence time that ultimately bounds the period over which signal can be acquired. While the native $T_2^*$ time is relatively short $\app$1ms, control techniques that can prolong the coherence time to $T_2$ or $T_{1\xr}$ can substantially boost image SNR. We will explore these aspects in a forthcoming manuscript.

Although making further comparison between both optical and MR modalities requires detailed consideration of the quantum efficiency and amplifier noise in both detectors, the discussion above already demonstrates that both modalities generally have complimentary advantages and are hence are highly amenable to combination. Considering photon losses alone, MR imaging provides a distinct advantage when compared to optical imaging: $\eta_m= {\displaystyle \prod_{i=1}^{3} \eta_{m,i}}=8.5\times10^{-8}$. This highlights that MR offers advantages for imaging particles that are buried under a scattering media.

%\begin{figure}[t]
%  \centering
%  \includegraphics[width=0.45\textwidth]{supplement_fig/snr_ratio-t.pdf}
%  \caption{\textbf{SNR ratio between optical and MR images.} We consider here the relative SNR between optical and MR modalities with depth of particles in tissue, assuming an initial (zero depth) SNR ratio $\xg_0=3$ as in our experiments. As is evident, the absence of optical scattering and attenuation loss in the MR modality can allow the latter to outperform optical imaging at relatively shallow particle depths. \I{Inset:} We consider here the critial depth $d_c$ at which $\xg=1$, i.e. the threshold at which MR imaging can outperform optical imaging for typical values of the scattering coefficient $R$.}
%\zfl{snr_ratio}
%\end{figure}

\subsection{Estimation of imaging SNR in scattering media}
\label{snr_section}
We compare the \I{relative} image SNR for optical and MR imaging modalities in scattering media. Defining the ratio of the two SNRs, $\gamma (d)$, where $d$ is the depth of the diamond particles. In our experiments (in the absence of tissue), $\gamma_0 \equiv \gamma(0)=\frac{\mathbf{SNR_{optical}}}{\mathbf{SNR_{MRI}}}=15$. If however, the particles are embedded in a scattering media with a depth of $d_0$, we expect an additional loss by a factor of $\R{exp} \{ \int_{0}^{d_0}- [\alpha(d)+\mu(d)]\cdot \R{d}d\}$ in optical imaging. At a critical depth $d_c=\frac{\log{\gamma(0)}}{\alpha_0+\mu_0 }$, the optical and MR image SNR become equal ($\gamma(d_c)=1$). This suggests that MR modality will produce better SNR for objects are beneath $d_c$. %In \zfr{snr_ratio}, we demonstrate the expected depth profile assuming an attenuation coefficient at 650 nm to be $\alpha_0$=0.1mm$^{-1}$. The critical depth $d_c$ is then typically under 1 mm, for realistic values of the scattering constant $R$.
Taking the attenuation and scattering coefficient used in Table \ref{3regime_table}, we have $d_c =$ 2.2 mm.

\subsection{Estimation of imaging resolution in scattering media}
\label{res_section}
\begin{figure}[h!]
\centering
\includegraphics[width=0.45\textwidth]{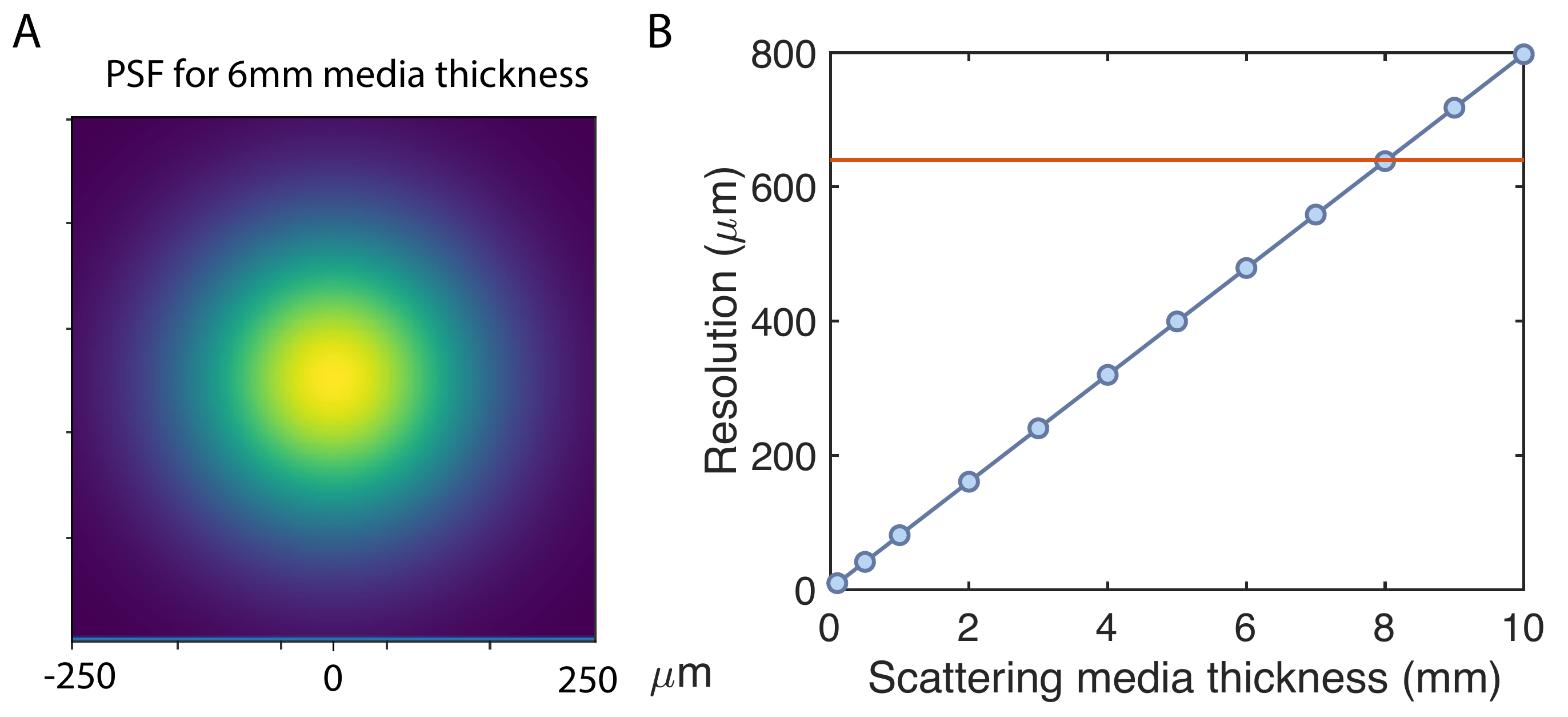}
\caption{\textbf{Image resolution in scattering media}. (A) Simulated wide-field PSF when the point source is scattered by a 6mm-thick media. (B) Relation between resolution and scattering media thickness. Blue dots denote the optical resolution. Red line denotes the MRI resolution corresponding to our experiments.}
\zfl{resolution}
\end{figure}

Besides SNR reduction, scattering blurs optical images as well, causing resolution reduction. In contrast, MR images are immune to optical scattering, with a resolution primarily being set by gradient strength and coherence time. Indeed, while optical imaging can easily achieve $\sim 1 \mu$m resolution, its performance degrades rapidly with increasing thickness of the scattering media. To illustrate this, we use multi-slice model~\cite{maiden2012ptychographic} to simulate the blur of a point source beneath a certain thickness of scattering media. We use a Wigner function $W(\vec{r},\vec{u})$ to represent the 4D light field, where $\vec{r}=(x,y)$ is the spatial coordinates, and $\vec{u}=(u_x,u_y)$ is the spatial frequency coordinates which is related to the propagation angle $\theta$ through $ \sin (\vec{\theta})=\lambda \vec{u} /n_r$. By assuming the scattering media is homogeneous and the spreading of angle $\vec{\theta}$ is Gaussian distributed, Wigner function $W(\vec{r},\vec{u})$ of the output field can be derived by summing up the propagation and scattering effect from each slice. Propagating through each slice has a tilting effect on $\vec{r}$ and the scattering due to each slice provides a Gaussian blur to $\vec{u}$~\cite{liu20153d}. Integrating over the frequency coordinates $\vec{u}$ to simulate the measured intensity of a wide-field microscope focusing at a point source through a scattering media of thickness $t$, we obtain,
\begin{equation}
\begin{aligned}
I(\vec{r}) & = \iint d^2 \vec{u} \,W(\vec{r},\vec{u}) \\
& =\iint_{(\frac{-n_r \theta}{\lambda} ,\frac{-n_r \theta}{\lambda})}^{(\frac{n_r \theta}{\lambda},\frac{n_r \theta}{\lambda})} d^2 \vec{u} \, \frac{n_r^2}{2 \pi \lambda^2 t^2 \sigma^2} \cdot \\
&    \exp{\left [\frac{n_r^2}{2 \lambda^2 t^2 \sigma^2} \left (\vec{r} - \frac{\lambda t}{n_r}\vec{u} \right )^2 \right ]}
\end{aligned}
\end{equation}
The Gaussian standard deviation $\sigma$ here quantifies the level of scattering. As an illustration in \zfr{resolution} we employ a typical value $\sigma =0.052$ corresponding to somatosensory cortex of a mouse brain~\cite{pegard2016compressive}. Assuming $\theta = 0.2527 $ rad representing angle of the acceptance of a $0.25$ NA objective, we calculate the FWHM of the blurred point spread function as resolution and plot the resolution versus media thickness in \zfr{resolution}B. Since we use homogeneous tissue assumption and $\sigma$ stays constant, we expect an approximately linear relationship between the blurriness and media thickness. Clearly, optical resolution is superior at shallow depths but MRI can outperform it when the tissue is more than $\sim$10 mm thick (\zfr{resolution}B). This results from comparison of the simulated optical resolution with the MRI resolution obtained in our experiments.

\section{Materials}

\subsection{Diamond particles}
The microcrystalline diamond powder involved in experiments in main Figure 1 has $\sim$40 mg mass and $\sim$200 $\mu$m size. The particles enriched with $\sim$1 ppm NV centers are from Element6, fabricated by high pressure high temperature (HPHT) growth. We emphasize that while these somewhat large particles were employed as a proof of concept demonstration, dual-mode imaging is currently viable even on smaller diamond nanoparticles. In fact, we have obtained hyperpolarization enhancements of $\lesssim$10 over 7T on commercially available (Adamas) 100nm fluorescent nanodiamonds~\cite{ajoy2020room}. The enhancement is limited primarily by material degradation stemming from radiation damage due to high fluence electrons employed to create the vacancy centers, as well as mechanical stresses incurred upon sample crushing. Recent work in~\cite{gierth2019high} has shown that annealing the diamond samples at ultra-high temperatures $\sim$1750 $^{\circ}$C can largely relieve this radiation-induced lattice disorder and enhance hyperpolarization levels by about an order of magnitude. These material advances along with the use of $\Cs$ enrichment portend signal gains that can allow translation of dual-mode imaging to nanodiamonds.

\subsection{$T_1$ relaxation in diamond}

\begin{figure}[t]
  \centering
  \includegraphics[width=0.48\textwidth]{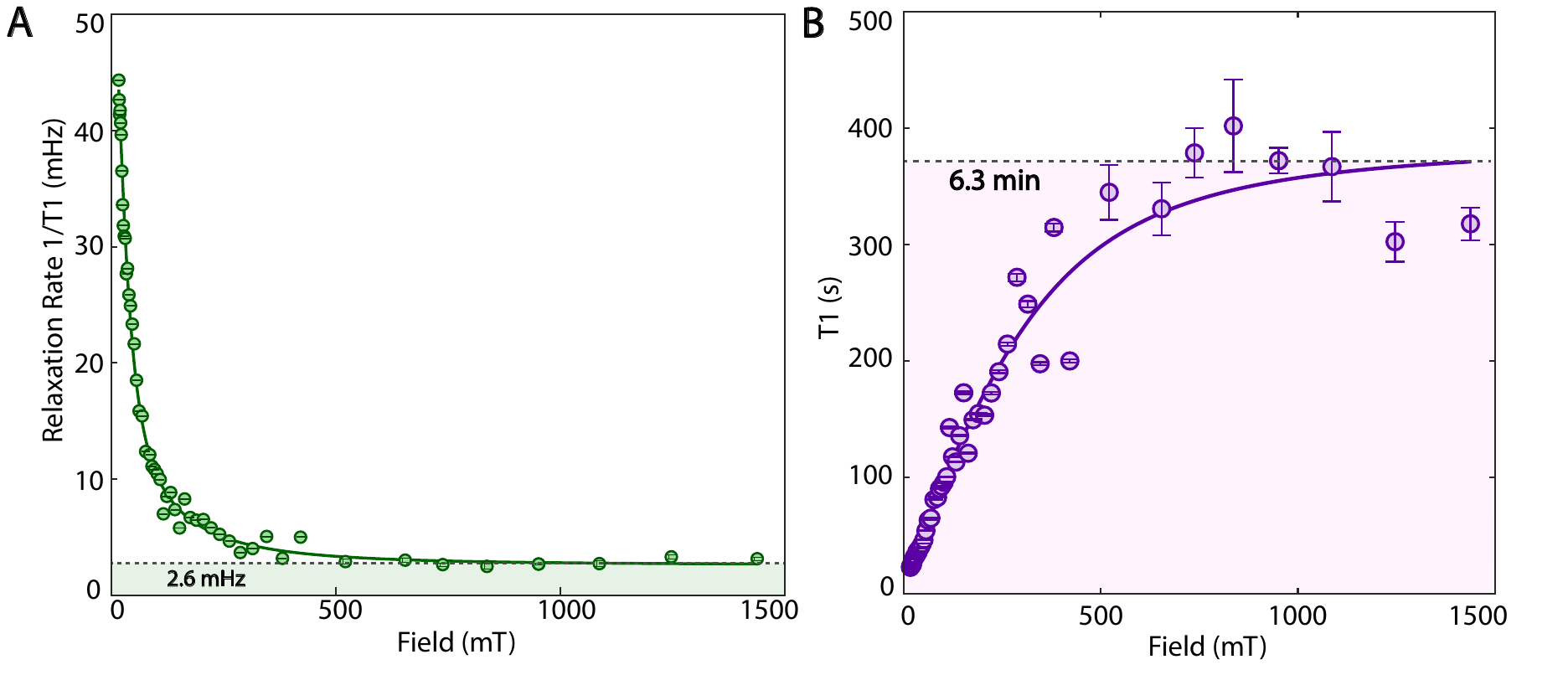}
  \caption{\textbf{$T_1$ lifetimes of $\Cs$ nuclei}, measured with 200$\mu$m diamond particles employed in this work. (A)$\Cs$ relaxation rate ($R_1(B) = 1/T_1$) mapped as a function of magnetic field, showing a sharp increase below $\sim$100mT. Solid line is a Lorentzian fit and error bars are calculated from mono-exponential fits (see \cite{Ajoy19relax}). (B) Corresponding relaxation times $T_1$, emphasizing that long relaxation times ( $>$6 min) are easily achievable even at modest fields. }
\zfl{t1}
\end{figure}

The several minute long $\Cs$ $T_1$ times of the diamond particles present advantages when they are employed as imaging agents~\cite{rej2015hyperpolarized}. In several respects the long $T_1$ times here are analogous to the long spin lifetimes of ${}^{29}$Si spins in silicon nanoparticles\cite{Cassidy2013} that made the latter systems highly attractive as MR imaging agents. In contrast to hyperpolarized silicon however, optically induced $\Cs$ hyperpolarization in diamond is replenishable and generated at room temperature and with much higher throughput.

We comment that the $T_1$ process in defect-center rich diamond particles primarily proceeds due to interaction of the $\Cs$ spins with lattice paramagnetic impurities. Ref. ~\cite{Ajoy19relax} has a detailed exposition on this subject, but for clarity we show in \zfr{t1} the field dependence of the $\Cs$ $T_1$ relaxation times of the particles here. In particular, the data illustrates sharp field-dependent relaxation profiles (see \zfr{t1}A) where the $T_1$ time drops sharply at low field. Intuitive understanding of the sharp profile is: the $\Cs$ Larmor frequency lies within the linewidth of the paramagnetic defect ESR spectrum (primarily P1 centers and radiation induced paramagnetic defects), engendering many-body flip-flops that can act as nuclear polarization relaxation channels. Fortunately however, the $T_1$ increases and saturates to multiple-minute long values beyond a modest a knee field $\sim$100mT. $T_1$ at this field is long enough to allow for simple polarization retention in permanent magnets.

\subsection{Hyperpolarization in $\Cs$ enriched diamond}

$\Cs$ enrichment provides a simple means to increase the effective number of $\Cs$ spins in the lattice at fixed sample volumes, and thus potentially increase MR image SNR. In single-crystal diamond samples produced by chemical vapor deposition (cvd) with controlled concentrations of $\Cs$ methane, we observe that the hyperpolarization enhancements per $\Cs$ nucleus can be approximately maintained to the same level up to an enrichment of about 10\%~\cite{Ajoy19relax}. %A detailed theoretical understanding to the limits to this observation, and its dependence on sample characteristics still remain out of reach. 
If translatable to diamond particles, this could allow increase of DNP signal by a factor of $\sim$10 (time savings of $\sim$100). We also observe a concomitant decrease in the $T_1$ relaxation times (and polarization buildup times) at higher enrichment~\cite{Ajoy19relax}. Overall these results point to order of magnitude SNR gains if HPHT based particulate diamond materials can be fabricated with high lattice $\Cs$ enrichment homogeneity. Efforts in this direction are underway in our laboratory.

\subsection{Biocompatibility}
Our focus of this work is to highlight the potential for dual-mode optical and MR imaging brought upon by optically hyperpolarizable agents. While NV centers in nanodiamonds are still the prototypical system of this class, there has been a recent spurt of activity of molecular systems with rare-earth or transition metal ions that share the same properties. Toxicity considerations remain an important consideration for all biomedical applications of such materials, but our goal was to highlight methodological advantages that might arise through the use of optical and MR modes, either simultaneously or in consort with each other. Here we provide a brief summary regarding biocompatibility of nanodiamonds. 

Diamond material is inert and chemically stable, and such properties could prevent its reaction inside body. Based on summarizing extensive toxicity studies in Table 2 in Ref.~\cite{turcheniuk2017biomedical}, Turcheniuk \textit{et al.} have drawn a general conclusion that nanodiamond derivatives of different origin and sizes do not have damages on basic functions of cells, organs, and animals in a reasonable range of concentrations. 

More specifically, low \textit{in-vitro} cytotoxicity has been shown in a couple of studies. For example, the toxicity of nanodiamonds was measured on human lung A549 epithelial cells and HFL-1 normal fibroblasts. The results showed that carboxyl-modified nanodiamonds with a particle size of 5 nm and 100 nm did not reduce the cell viability or affect the protein expression at the concentration of 0.1-100 $\mu$g/ml, ~\cite{liu2007biocompatible}.

\textit{In-vivo} toxicity studies also found no significant impact of nanodiamonds on animal health~\cite{yuan2010pulmonary}. The pulmonary toxicity of the nanodiamonds was studied in experiments where they are administered in mice through intratracheal instillation. The work found no noticeable adverse effects in the lungs from the nanodiamonds within the period according to the histopathological and ultrastructural investigations. The results also indicated that nanodiamonds can be extruded through engulfment by lung macrophages. From the studies, we can see that the respiratory toxicity of nanodiamonds, in a wide variety of dose range, is less than that of CNTs and other carbon nanomaterials that are extensively studied in the previous literature~\cite{lam2004pulmonary,muller2005respiratory}.

Long term studies have been done by replacing water in the mice's diet with nanodiamond hydrosols for 3 to 6 months~\cite{schrand2009nanodiamond}. Neither death nor abnormal in growth or the internal organ (liver, lungs, heart, kidneys, and pancreas) caused by nanodiamonds was found in the results comparing with the control mice indicated. The same study has also shown that there are no effects on the mouse's reproductive ability. No inflammatory symptoms in the mice has been observed after subcutaneous exposure to nanodiamonds for 3 months~\cite{puzyr2007nanodiamonds}. Furthermore, no immune responses was observed with 10 days after intraperitoneal injection of nanodiamonds~\cite{bakowicz2002wide}. It was also found that after filtration to remove large size particles, the uptake of nanodiamonds was significantly reduced in the lung, the spleen and the liver~\cite{rojas2011biodistribution}. Long period observation (5 month) confirmed no toxicity for dosages as high as 75 mg/kg body weight of nanodiamonds by histopathological analysis of various tissues and organs in mouse models over long period of time~\cite{vaijayanthimala2012long}.

Although diamond nanoparticles tend to remain in body for longer time, it has been found that in a mouse model, the diamond nanoparticles were excreted through urine~\cite{rojas2011biodistribution}. Specially, a gradual decrease of concentrations in lung, spleen, and liver was observed, and an increase in bladder by PET imaging of $^{18}$F labeled nanodiamonds was shown. And 2 hours after administration of nanodiamonds in rats, higher concentration was found in urine than organs.

Given the low toxicity results in both \textit{in-vitro} and \textit{in-vivo} studies as well as the elimination pathway through urine, nanodiamonds have been proposed for cancer therapy, drug delivery, antimicrobial agents as well as biomedical imaging. For example, diamond as a drug delivery carrier has been reported in tissue level~\cite{zhang2011multimodal,xi2014convection} and mouse models~\cite{chow2011nanodiamond}. The diamond-amoxicillin complex in Gutta percha matrix is now in clinical trial as a root canal tooth implant~\cite{ho2020nanodiamond}.

\section{Methods}

\subsection{$\Cs$ Hyperpolarized MRI experimental setup}

\begin{figure}[t]
  \centering
  \includegraphics[width=0.45\textwidth]{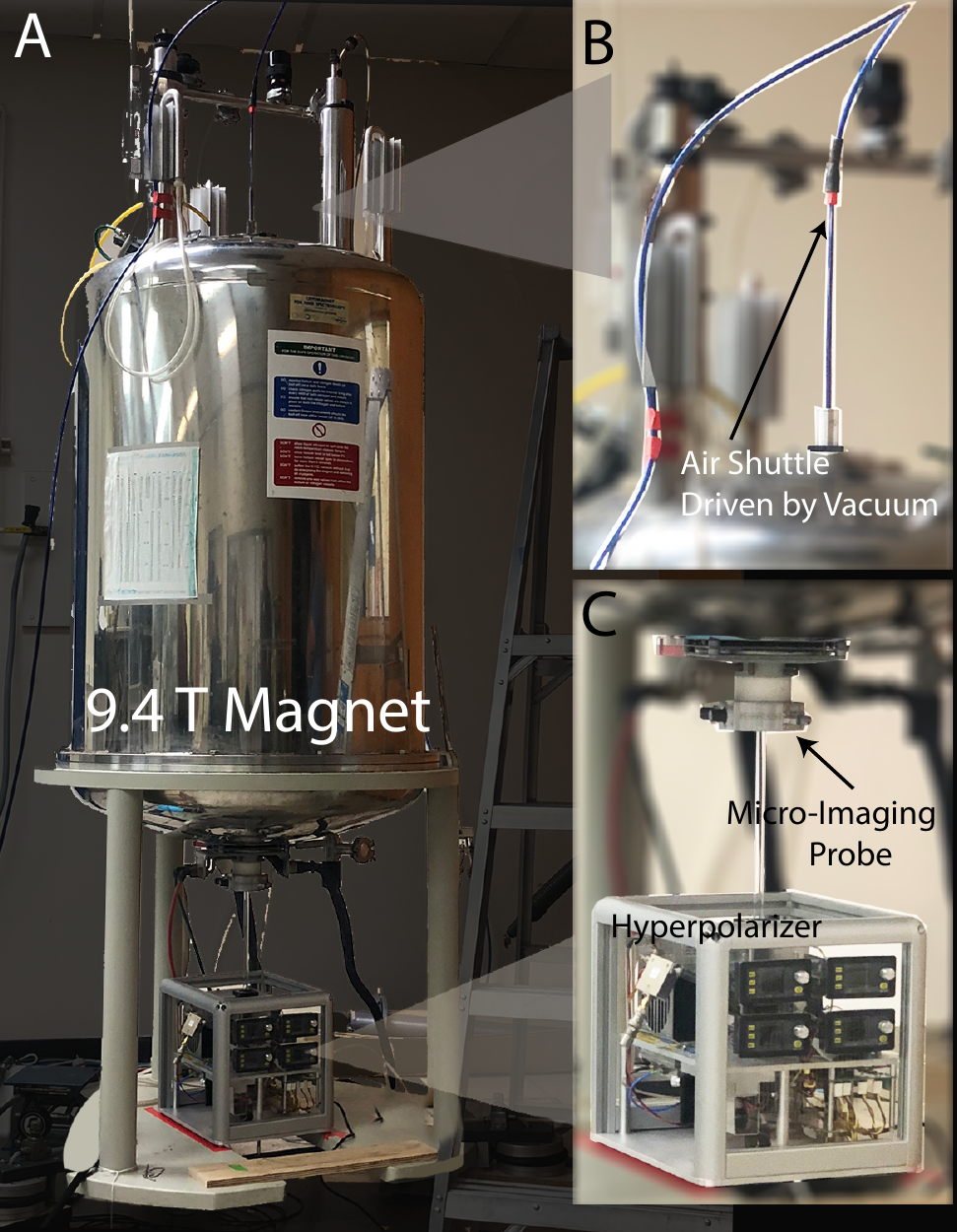}
  \caption{\textbf{Hyperpolarization experiment setup} consists of (A) a 400MHz (9.4T) Bruker DRX system fitted with a micro-imaging probe (10 mm $\Hs$/$\Cs$ volume coil) for MRI acquisition, and {(B)} an air shuttle based field-cycling system that transports the sample from low to high field. {(C)} Polarization is generated in a portable polarizer device placed at the bottom of the magnet.}
\zfl{setup}
\end{figure}

We now detail the setup employed for diamond hyperpolarization and $\Cs$ MR imaging (\zfr{setup}A). It consists of a pneumatic field-cycling device, a wide-bore 9.4T superconducting magnet, and a miniaturized hyperpolarizer (Hypercube,~\cite{ajoy2020room}). The field-cycling device (\zfr{setup}B) enables rapid sample transfer from low field (40 mT) to the 9.4 T detection field, within which a 10 mm $\Hs$/$\Cs$ volume coil is installed (\zfr{setup}C). It is composed of a quartz channel in which the sample is transported, a concave-shaped stopper at the bottom end of the channel, and is driven by a pump that generates the requisite vacuum to drive the transfer. The overall transport time into the magnet is under 1s, short compared to the $\Cs$ $T_1$ times. MR imaging was conducted with a Bruker DRX system and custom programmed `FLASH' pulse sequence.

The device  (\zfr{setup}C) delivering hyperpolarized diamond particles is small in size ($\sim$25 cm edge) compared to traditional DNP apparatuses (see Ref. ~\cite{ajoy2020room} for a detailed description). The self-contained unit encapsulates components for laser excitation, and MW irradiation as well as an electromagnet for field fine-tuning. We employ a a 1W 520nm diode laser (Lasertack PD-01289) and passes through an aspheric lens and a set of anamorphic prisms to form a 4 mm diameter beam. The beam was routed by two mirrors to illuminate the sample from the bottom. MW irradiation that drives polarization transfer is generated by three voltage controlled oscillator (VCO) sources (Minicircuits ZX95-3800A+). For frequency sweeps, the VCOs are driven by phase shifted triangle waves from a home-built PIC microprocessor (PIC30F2020)  driven quad ramp generator. A typical sweep frequency at 200Hz, approximately determined by the NV $T_{1e}$ lifetime~\cite{ajoy2020room}, is used to sweep through NV powder spectrum. The on board Hall probes measure the vector magnetic field experienced by the diamond particles, and an algorithm that determines the position of the extrema of the NV powder pattern to sweep over. 
The VCO signals are then power combined and amplified by a 16 W amplifier, and delivered with a loop antenna. The broadband antenna has substantial reflection and consequent power loss, which we estimate to be greater than 90\%. We typically employ a polarizing field $\sim$38 mT, where hyperpolarization enhancements are optimal~\cite{ajoy2020room}. 

\subsection{Evaluation of SAR and optical power}

In order to characterize the MW induced heating in potential future deployment of hyperpolarization in in-vivo settings, we detail the evaluation of the Specific Absorption Rate (SAR)~\cite{Goodwill10} with parameters in our experiments. We suppose the MW field implemented is $B_{1}$, causing heat generation. Considering this field applied in a cylindrical space between two split coils, we derive the eddy current and the heat generated by the current and obtain:
\begin{equation}
\textrm{SAR} = \frac{\sigma \pi^2 f^2 B_{1}^2 D^2}{8 \rho}\:,
\end{equation}
where $\sigma$ and $\rho$ are respectively the conductivity and density of the media; $\sigma = $0.57 S/M\cite{zurbuchen2017determination} and $\rho = $1079 kg/m$^3$ for human tissue. Assuming $f \approx$ 2.9 GHz is the MW frequency, and $D=$14 mm is the diameter of the split coil, we obtain a SAR = 1.1$\times10^{12}B_{1}^2$ W/(kg$\cdot$T$^2$) or 1.1$\times10^{4}B_{1}^2$ W/(kg$\cdot$G$^2$) as a function of the $B_1$ field employed. A direct measure of the electron Rabi frequency is somewhat challenging in our setup, due to MW field inhomogeneity and specifics of the MW delivery platform employed. However we can provide an estimate from a theoretical evaluation of the Landau-Zener hyperpolarization mechanism. Using the estimated $\xO\lesssim$200 kHz\cite{zangara2019dynamics}, We calculate a SAR of $\sim$14 W/kg. The heating issue needs to be addressed while maintaining enough signal. We note importantly that the hyperpolarization enhancements fall approximately logarithmic with MW power beyond a threshold~\cite{ajoy2018ori}, and we can have $\textrm{SNR} \propto \log (B_1)$ from data shown in \zfr{siganl_rabi}. In such regime, the relationship between SAR and SNR is: $\textrm{SAR} \propto B_{1}^2 \propto \textrm{exp}(2\textrm{SNR})$. This indicates that it is possible to reduce the SAR by lowering the MW power with a benign loss in the hyperpolarization signal ($\sim$47 \% loss) to reach an FDA SAR limit (4 W/kg). Potential SNR improvement through better engineered materials or higher filling factors can compensate such loss and allow further SAR reduction.

\begin{figure}[t]
  \centering
  \includegraphics[width=0.30\textwidth]{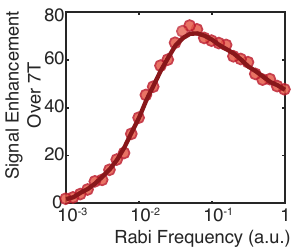}
  \caption{\textbf{Hyperpolarization signal as a function of Rabi frequency.} (Data in~\cite{ajoy2018ori}) Hyperpolarization signal shows approximately logarithmical dependence on Rabi frequency when MW power is smaller than a threshold.}
\zfl{siganl_rabi}
\end{figure}

Let us comment on optical power as well. The MRI demonstrated in main Figure 1F employed a laser power density of $\sim$80 mW/mm$^2$. While this is elevated above levels suitable for in-vivo operation ($\sim$4.7 mW/mm$^2$~\cite{Denton06}), this is primarily on account of the large mass ($\sim$40 mg) of the particles in this experiment. We anticipate that laser power will not be restrictive in preclinical imaging settings since far lower diamond masses are typically employed in the relevant applications.

\begin{figure}[t]
  \centering
  \includegraphics[width=0.45\textwidth]{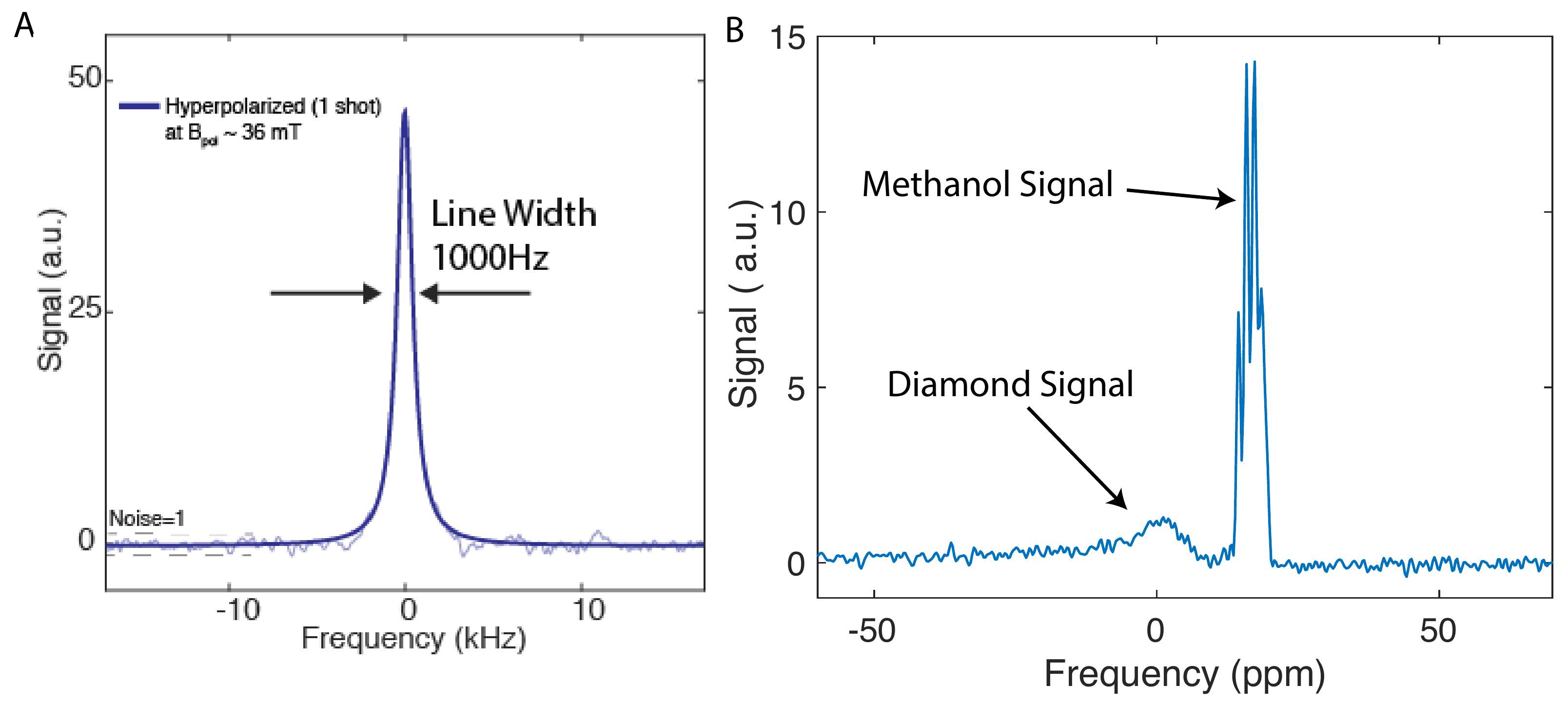}
  \caption{\textbf{$\Cs$ NMR spectrum} of diamond and diamond/methanol mixture relevant to the MR background suppression experiments in Figure. 3 of the main paper. (A) Typical hyperpolarized $\Cs$ spectrum of diamond with line width $\approx$1kHz (data from~\cite{ajoy2020room}). (B) Diamond particles hyperpolarized in a $\Cs$ methanol mixture, showing that the two spectra overlap with each other. Note that the in the MR imaging sequences, the pulse bandwidth is 213kHz, and sampling bandwidth is 900 kHz, which are inadequate to distinguish the two sources.}
  %Instead we employ controlled sign reversal on hyperpolarization signal to completely suppress the [$\Cs$]-methanol background.}
\zfl{spec}
\end{figure}

\begin{figure}[t]
  \centering
  \includegraphics[width=0.4\textwidth]{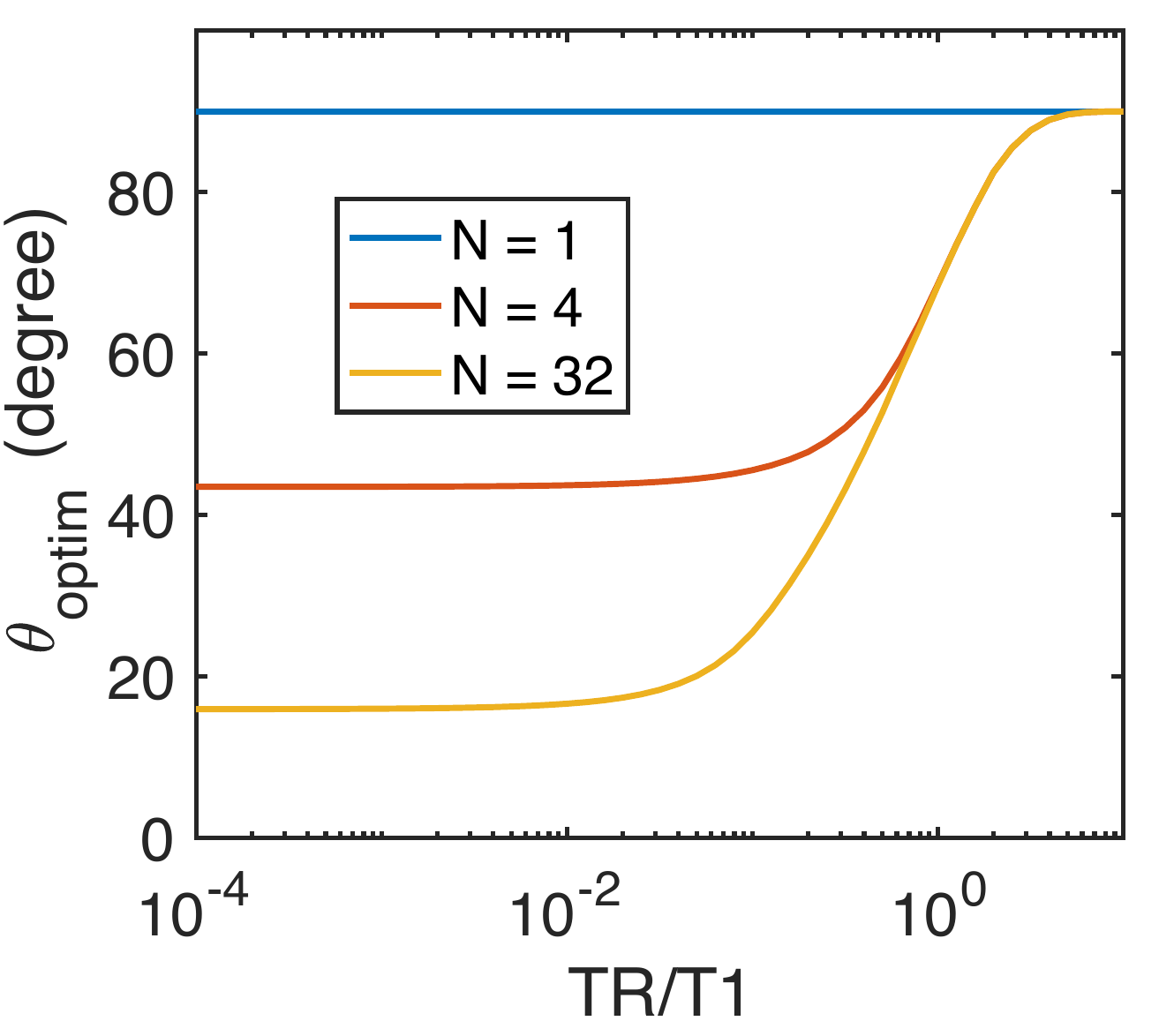}
  \caption{\textbf{Optimization of flip angle} $\xt$ employed for MR imaging. Here we consider simulated traces for the case of image consisting of N = 1, 4, 32 slices, and plot the flip angle $\xt_{\R{opt}}$ that maximizes the total transverse magnetization $S_{\text {cumulative}}$ over the entire imaging sequence. Our experiment corresponds to the regime where TR/T1 $\sim 10^{-3} $, indicating the optimal flip angle for $N=32$ employed in the images in main Figure. 1F is $\sim$15$^{\circ}$.}
\zfl{flip_angle}
\end{figure}

\begin{figure}[t]
  \centering
  \includegraphics[width=0.45\textwidth]{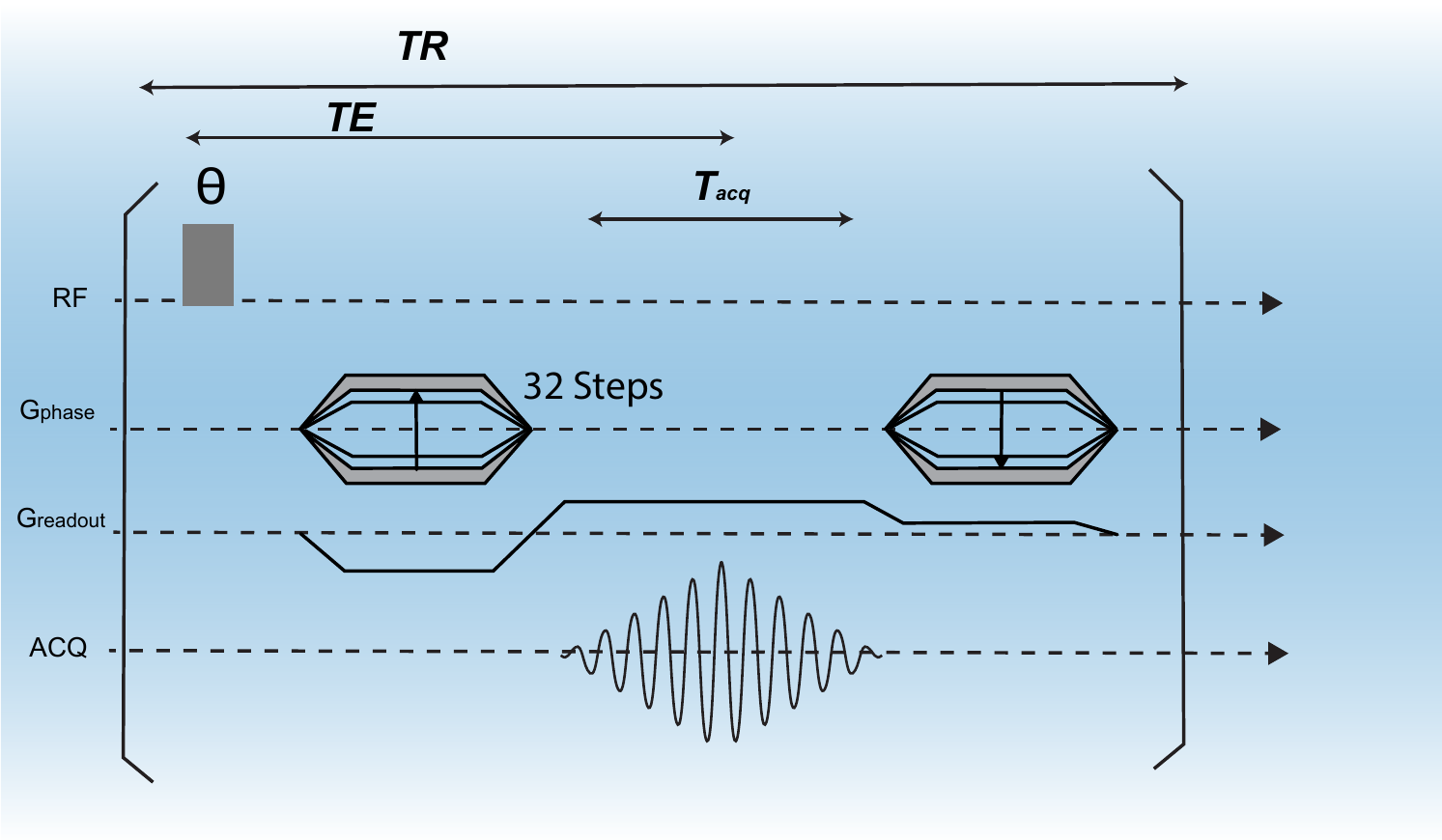}
  \caption{\textbf{MR imaging sequence.} A variant of FLASH is used in the MRI experiments in this paper. Here echo time TE = 0.5 ms, repetition time TR = 6 ms, acquisition time T$_{acq}$ = 0.36 ms, and we employed 32 phase encoding and 32 frequency encoding steps.}
\zfl{imaging_sequence}
\end{figure}

\subsection{Magnetic resonance imaging}
We now consider details of methodology employed to produce the MR images in Figure. 1F of the main paper. The short $T_2^{\star}\sim$1 ms of the diamond $\Cs$ nuclei (see \zfr{spec}A), as typical in any solid state system~\cite{Frey12}, present technical challenges for MR imaging. We optimize our gradient echo imaging pulse sequences (see \zfr{imaging_sequence}) to account for fast spin dephasing. Specifically, the width of RF pulses and gradient lobes are minimized to create as short an echo time as possible (TE = 0.5 ms), and fast gradient switching (200 $\mu$s) is employed for phase encoding. Moreover, since we are only interested here in 2D projection images, the slice selection gradient $G_{slice}$ was removed in our variant of FLASH. This removes transient interference with the subsequent phase encoding gradient and reduces distortions during the short RF pulse. Gradient spoilers were employed on both the phase encoding and readout channels at the end of the sequence to destroy the residual spin coherence and has minimal impact on the subsequent pulse cycle. Finally, taking advantage of the substantial initial polarization, we use small flip angles ($\xt$ in \zfr{imaging_sequence}) to optimally retain polarization over the entire imaging sequence, distributing the magnetization in each of the phase encoding steps in the most efficient manner. 

We now analyze the optimal flip angle employed for imaging. Due to the limitation of the instrumentation, we can only have a constant flip angle $\xt$ in all the repetitions. If $M_n$ represents the total magnetization at the beginning of the $n$th repetition of the imaging sequence, and $M_0$ represents the thermal magnetization, we can compute a recursive relationship between $M_n$ and $M_{n-1}$, as well as the transverse magnetization $M_{x, n}$ as :
\begin{equation}
\left\{\begin{array}{c}{M_{n}=\left(M_{n-1} \cos \theta-M_{0}\right) e^{-\frac{\R{TR}}{T_{1}}}+M_{0}}
 \\ {M_{x, n}=M_{n} \sin \theta}\end{array}\right.
\end{equation}
where $\R{TR}$ is the repetition time, and $\theta$ the flip angle used to generate transverse magnetization. We consider the cumulative transverse magnetization $S_{\R {cumulative}}=\sum_{n=1}^{N} M_{x, n}$. Assuming that $M_1\gg M_0$ is self-evident under hyperpolarization, we can optimize $\theta$ to maximize $S_{\text {cumulative }}$ (see \zfr{flip_angle}). For a total of $N=$32 repetitions, $\xt_{\R{optim}}=$15$^{\circ}$ corresponding to a 6$\mu$s pulse width under 2dB RF attenuation. 

The imaging sequence implemented in our experiments is detailed in \zfr{imaging_sequence}. We employed hard pulses with bandwidth 213kHz, which is sufficient to cover the entire $\Cs$ spectrum. The small flip angle sequence is repeated 32 times to obtain 32 lines in phase encoding dimension, and each phase encoding step includes 32 points in frequency encoding dimension. The sampling dwell time is 0.011ms, giving a sampling bandwidth of $\app$ 90kHz. Sampling in phase encoding dimension proceeds in a centric fashion, k=$\{0, -1, +1,\cdots -16\}$, so as to place the largest signal at the center of k-space, and a FFT finally produces the MR image. In our experiment, a typical spatial resolution is $0.64$ mm $\times 0.64$ mm, which after zero-filling and smoothing results in a square pixel size (main Figure. 1F) of $0.16$ mm. We note that since in the absence of slice selection, any image we obtain is a 2D projection along z-axis.

Let us comment on approaches to extend the diamond particle imaging to 3D. The short $T_2^*$ of $\Cs$ in diamond that prevents longer acquisition is the major challenge in the cases of 3D imaging. Refocusing sequences such as magic echo sequence~\cite{takegoshi1985magic} or quadratic echo sequence~\cite{frey2012phosphorus} can prolong the coherence time during which imaging takes place. Ultrashort TE, or UTE sequences~\cite{chang2015ute} is another approach where gradient on both phase encoding and frequency encoding dimensions can be applied while the acquisition channel opens right after RF excitation. Such sequences can allow us to either implement slice selection or phase encode the information along the third dimension.

\subsection{Optical imaging}
We describe the details of optical imaging in this section. Optical imaging of diamond particle fluorescence in Figure. 1E employed four laser-diodes (Lasertack PD- 01289), coupled with optical fibers (Thorlabs M35L01) to uniformly irradiate the sample from four orthogonal directions. Note that, hyperpolarization is conducted with one laser. The fluorescence is filtered by a 630 nm long pass filter (Thorlabs FGL630) before light collection, either by an avalanche photodiode (APD) (Thorlabs APD410) or a CMOS camera (DCC1645C) equipped with an objective lens (MVL5M23). The APD output allows quantification of the overall sample fluorescence, and was recorded by a data acquisition card (NI USB 6229). Such APD data is presented, for instance, in Figure 2A of the main paper. The optical images and time elapsed data for background suppression are instead obtained from the CMOS camera, employing a 10Hz frame rate.

Adding a third dimension to optical imaging may require a confocal-type of setup, and scattering can largely deteriorate the quality of optical images. We envision that the dual-mode imaging capability can overcome the challenge by augmenting the optical images with three dimensional MRI.

\subsection{Photon counts estimation}
We estimate the overall photon counts in the optical imaging via the a Thorlabs APD410 photo diode. Without changing the imaging setup, we measured a voltage from the APD $V_{out} =3$V in front of the camera. Referring to the manual~\cite{apd410}, we have responsivity $R_{M}$(637 nm) = 33 A/W, and the corresponding incident light power per volt of electrical output $P_{icd}/V_{out}$=1.21 $\times 10^{-7}$W/V. We estimate the integrated photon counts in a typical image (main Figure 1E) to be $12 \times 10^{11} $ counts/s.

\subsection{Optical and MR image SNR estimation}
We now detail methods employed to estimate the image SNRs in the optical and MR images in Figure. 1 E, F of the main paper. We manually select a circular region of interest (ROI) to measure the mean signal value in the ``object region'', and move the same ROI away from the object to a background region to measure the noise level. The SNR is measured 10 times at different locations on both images, and an average is taken to quantify image SNR. We thus obtain a normalized one-shot SNR of 1.05 for the MR image in Figure. 1F. For optical image (Figure. 1E), the measured SNR is 12.0 in 0.1s of acquisition. In order to draw a fair comparison, we rescale this SNR to 3.0 for the same optical power employed in MRI. 

We then normalize the SNR against acquisition time. We obtain an optical SNR of 9.5$\sqrt{\textrm{Hz}}$, and MRI of 0.63$\sqrt{\textrm{Hz}}$ (with consideration of 40s dead time to hyperpolarize the sample).
%This corresponds to overall optical and MR SNRs of 30$/\sq{\R{Hz}}$ and 0.026$/\sq{\R{Hz}}$ in Figure. 1 of the main paper, where we have included the dead time (40s) involved in the hyperpolarization process. 
We note that MR SNR can be substantially improved beyond this value by several factors, including $\Cs$ enrichment, improvements in filling factor, material optimization~\cite{gierth2019high}, as well as carefully designed pulse sequence.
\begin{figure}[t]
  \centering
  \includegraphics[width=0.45\textwidth]{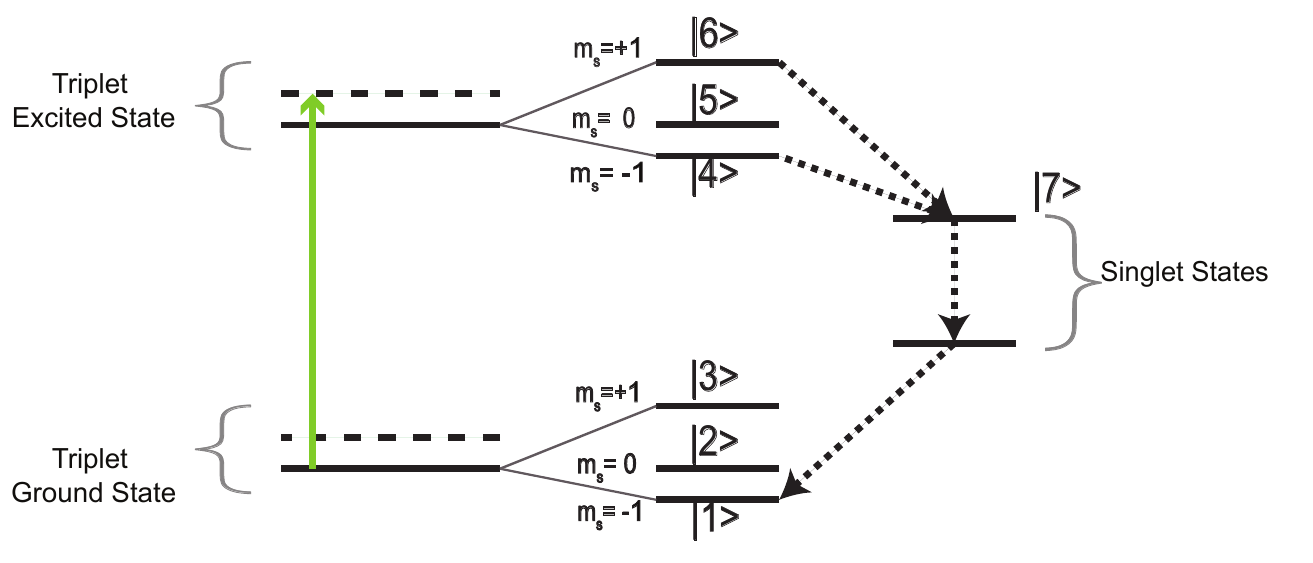}
  \caption{\textbf{NV energy level structure} simplified to a 7-level model. The NV- centers are electronic spin-1 with ground and excited triplet states, as well as a ``dark" singlet state. The m$_s$ = $\pm$ 1 states degenerate at zero field and split into two states when an external magnetic field is applied.} %At the low fields we operate under, the dominant term in the Hamiltonian is set by the zero-field splitting (along the N-to-V axis).}
\zfl{NV_energy-level}
\end{figure}

\begin{figure}[t]
  \centering
  \includegraphics[width=0.4\textwidth]{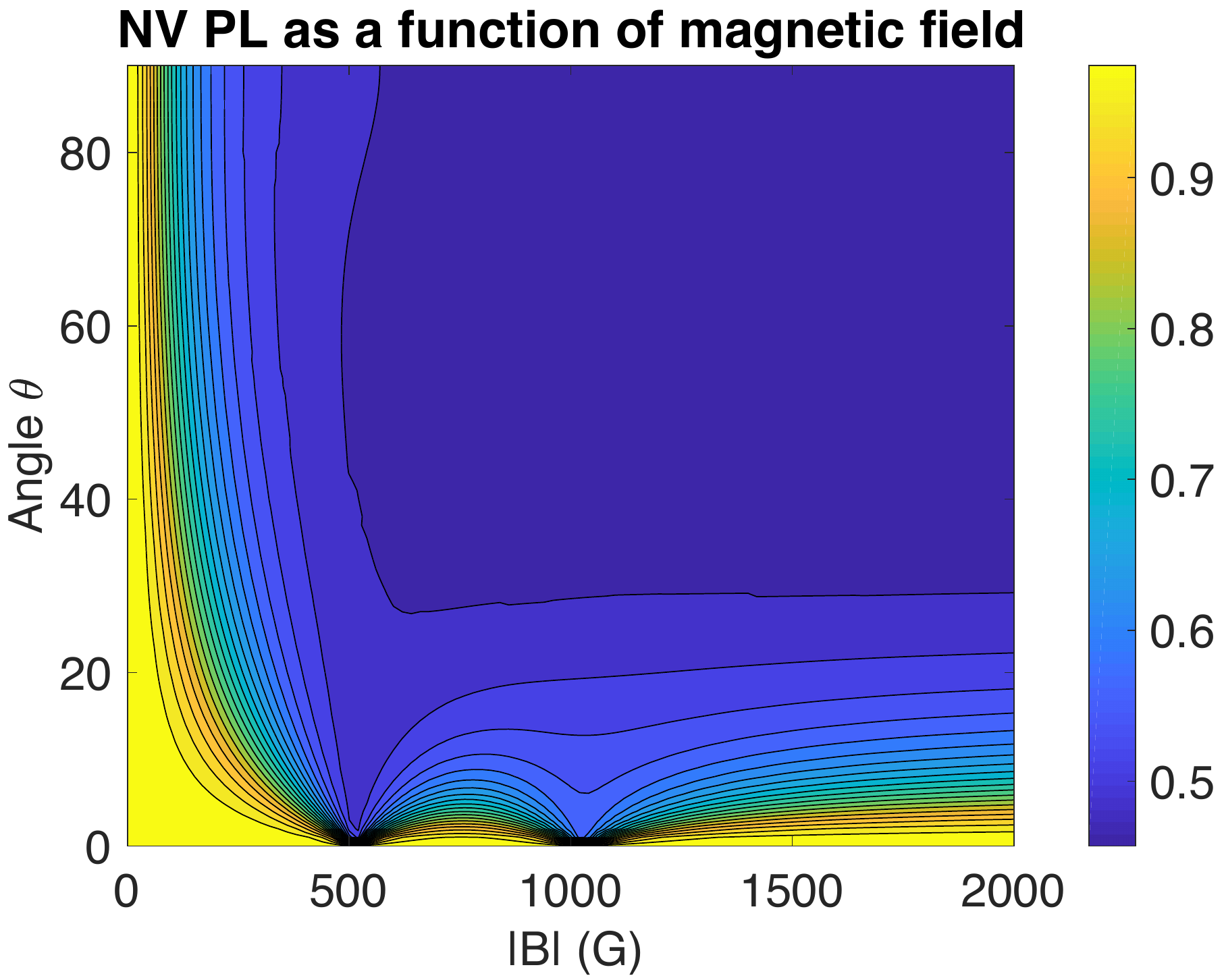}
  \caption{\textbf{NV photoluminescence} as a function of external magnetic field and the angle between magnetic field and the N-V axis. This is simulated from a 7-energy-level model of NV state kinetics assuming no charge interconversion between NV charge states. Highest contrast appears close to excited and ground state anti-crossing at 500G and 1000G.}
\zfl{NV_FL}
\end{figure}

\subsection{Pulsed field fluorescence modulation}
The magnetic field dependence of NV fluorescence enables amplitude modulation of optical images (main Figure 2A, B). We simulate this effect using the method in a pioneering work in Ref.~\cite{tetienne12} but include a detailed description here for completeness. When an external magnetic field is not perfectly aligned with the N-V axis, it lifts the degeneracy of the $m_s=\pm 1$ levels and mixes the $m_s=\pm 1$ and $m_s$ = 0 states. This effect can be simulated a 7-level-model of the NV- center (\zfr{NV_energy-level}), where the singlet dark states are considered as one state $|7\rangle$. Upon application of a static magnetic field ${\mathbf{B}_{ext}}$, the eigenstates {$|i\rangle$} of the system can be expressed as linear combinations of the zero-field eigenstates \(|i\rangle=\sum_{j=1}^{7} \alpha_{i j}(\mathbf{B}_{ext})\left|j^{0}\right\rangle\). The ground state and excited state Hamiltonian can be written as:
\begin{equation}
\begin{aligned}
 \mathcal{H}_{\mathrm{gs}}=h D_{\mathrm{gs}} S_{z}^{2}+g \mu_{\mathrm{B}} \mathbf{B}_{ext} \cdot \mathbf{S} \\
  \mathcal{H}_{\mathrm{es}}=h D_{\mathrm{es}} S_{z}^{2}+g \mu_{\mathrm{B}} \mathbf{B}_{ext} \cdot \mathbf{S}
 \end{aligned}
\end{equation}
 
where $D_{\mathrm{gs}}$ and $D_{\mathrm{es}}$ are zero field splitting of the ground state and the excited state. We diagonalize the Hamiltonian to estimate $\{\alpha_{i j}(\mathbf{B}_{ext})\}$. Following Ref.~\cite{tetienne12}, the dynamics of population in the different states can be evaluated as:
\begin{equation}
\frac{\mathrm{d} n_{i}}{\mathrm{d} t}=\sum_{j=1}^{7}\left(k_{j i} n_{j}-k_{i j} n_{i}\right)
\end{equation}
where $n_i$ is the population of the $\ket{i}$ state, and $k_{ij}$ denotes the kinetic transition rate between state $\ket{i}$ and $\ket{j}$. %Solving the system Hamiltonian, we obtained $k_{ij}$.
The transition rates $\{k_{i j}(\mathbf{B}_{ext})\}$ are associated with the zero-field transition rates $\{k_{p q}^{0}\}$ through the transformation \(k_{i j}(\mathbf{B}_{ext})=\sum_{p=1}^{7} \sum_{q=1}^{7}\left|\alpha_{i p}\right|^{2}\left|\alpha_{j q}\right|^{2} k_{p q}^{0}\). We employ experimentally measured values of $\{k_{p q}^{0}\}$ from Ref.~\cite{tetienne12}. Solving the steady state of the kinetic equations, we obtain steady state photoluminescence $\mathcal{R}$ as a function of applied ${\mathbf{B}_{ext}}$:
\begin{equation}
\mathcal{R}(\mathbf{B}_{ext}) = \sum_{i=4}^{6} \sum_{j=1}^{3} k_{i j}(\mathbf{B}_{ext}) n_{i}
\end{equation}

Evaluating $\mathcal{R}(\mathbf{B}_{ext})$ for different magnetic field strengths and angles with respect to the N-V axis, we plot a fluorescence map in \zfr{NV_FL}. Considering now a randomly oriented powdered diamond sample with sufficiently high mass such that all the angles $\theta$ between $\mathbf{B}_{ext}$ and N-V axis are uniform, the overall fluorescence as a function of applied magnetic field is:
\begin{equation}
D(|\mathbf{B}_{ext})|) = \frac{\int_0^{\pi} \mathcal{R}(|\mathbf{B}_{ext}|, \theta) \sin(\theta) d\theta}{\int_0^{\pi} \sin(\theta) d\theta}
\end{equation}
The result is shown in Figure. 2A of the main paper. We note that while this simple model points to an attainable contrast approaching $\sim$40\% for a field strength $\sim$40 mT, we experimentally obtain a lower contrast $\sim$ 10\%. We speculate that the discrepancy stems from NV charge state interconversion and light scattering, both of which are not accounted for in this model, and are challenging to estimate with precision in our experimental settings.

\subsection{Dual-mode background suppression}

\begin{figure}[t]
  \centering
  \includegraphics[width=0.4\textwidth]{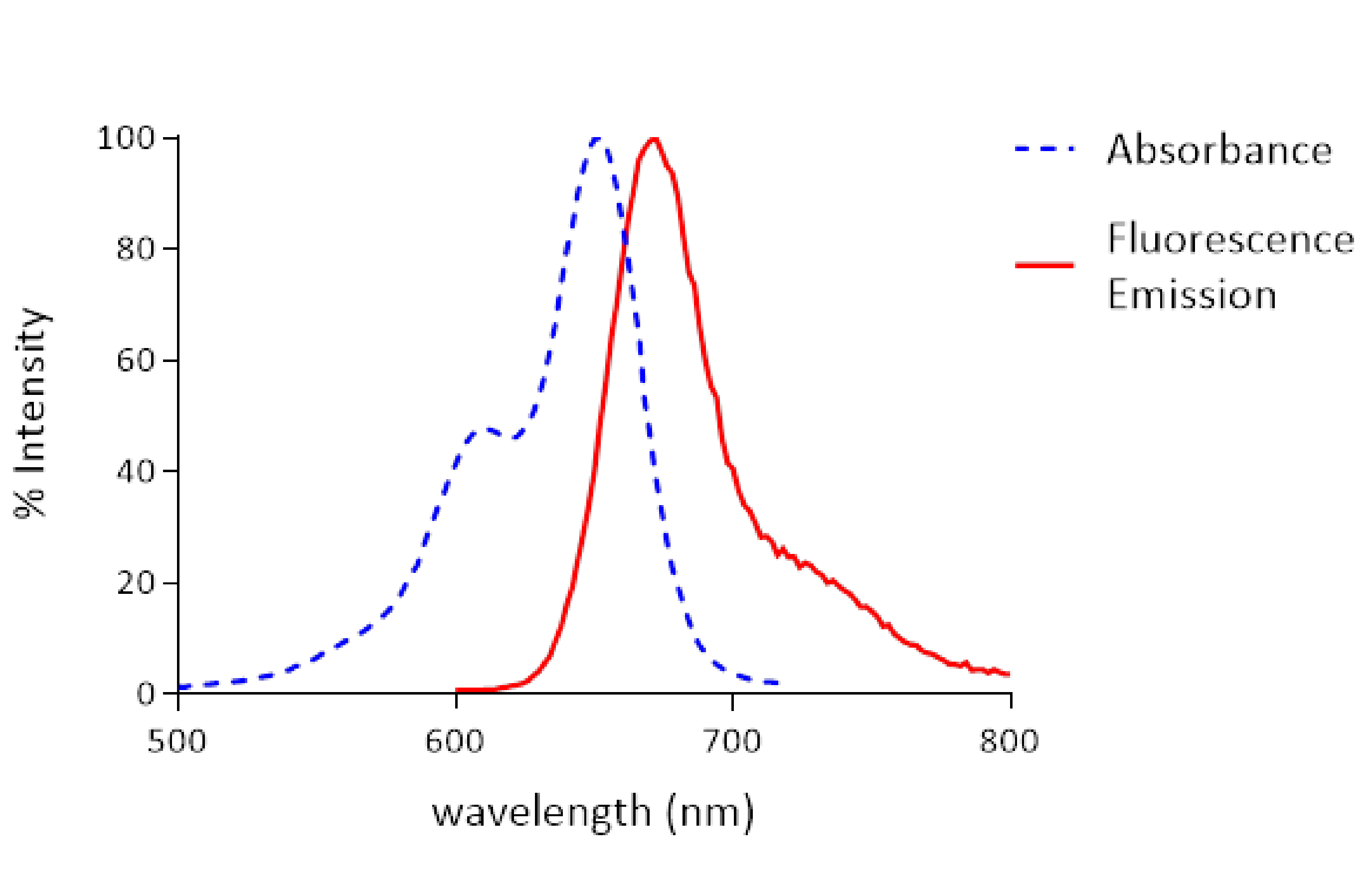}
  \caption{\textbf{Optical spectrum of the Alexa 647 dye} as a background employed in the experiments in Figure 3C of the main paper (adapted from datasheet~\cite{alexa647}). It is evident that the fluorescence emission of overlaps strongly with NV emission spectrum, making the diamond particles indistinguishable in the background but recoverable through lock-in modulation under a pulsed magnetic field.}
\zfl{optical_spec}
\end{figure}

Modulation of the images in both the optical and MR modalities allows effective suppression of background signals. We now provide more details about the experimental demonstration in Figure 3 of the main paper. For the optical images where  Alexa 647 overlapping with NV centers on the optical spectrum (\zfr{optical_spec}) is utilized as an artificial background, we employ software lock-in suppression. In such method, we record a 100s time lapse at 10 fps when modulating the signal of the entire image with a $\approx$400 G amplitude, 0.1 Hz square wave field applied to the sample. Due to the mechanism outlined above, only fluorescence from the diamond particles are modulated by the field change, and a peak at 0.1 Hz will present for these pixels in the Fourier domain, allowing the lock-in reconstructed wide-field background-free image is shown in Figure 3E. Although the images here constitute a proof of concept demonstration and the suppressed background is $\sim$2 times of the signal, the robust and non-blinking modulation from the particles engender background suppression factors that can be as high as 2 orders of magnitude~\cite{bumb14}.

Common mode rejection is used in $\Cs$ MRI for background suppression, exploiting the unique ability of our hyperpolarization method to entirely reverse the $\Cs$ polarization on-demand. We chose a background consisting of [$\Cs$]-methanol, since its spectrum (shown in \zfr{spec}B) appears very close to that of $\Cs$ in bulk diamond. The background suppressed is $\sim$5 times of the signal in Figure 5D. While this constitutes a proof a concept, we emphasize that the suppression factors can exceed two orders of magnitude~\cite{ajoy2020room}.

\subsection{Comparison with $^1$H relaxation contrast imaging}

$^1$H imaging with conventional exogenous contrast agents has been extensively used in MRI studies~\cite{merbach2013chemistry}. In our work, we propose to image the nanodiamond object directly in contrast to some contrast imaging techniques relying on imaging the relative difference of relaxation time of the objects. It is different from 1H contrast imaging in two aspects: (i) nuclei species: 13C imaging presents relatively smaller SNR but much lower background than 1H imaging. (ii) contrast mechanism: our hyperpolarization contrast is usually higher than relaxation (T$_1$ or T$_2$) contrast.

In order to compare two approaches, we can define a contrast-to-noise ratio (CNR) to quantify the image quality: $\mathrm{CNR} = \frac{\mathrm{signal \:difference}}{\mathrm{noise}}$. Assuming the signal of the object and background is $s_o$ and $s_b$ respectively, and the noise is $n$, we can rewrite CNR as:
\begin{equation}
\mathrm{CNR}=\frac{s_{o}-s_{b}}{n}=\frac{s_{o}}{n} \frac{s_{o}-s_{b}}{s_{0}}=\operatorname{SNR} \frac{s_{o}-s_{b}}{s_{o}}
\end{equation}

For the first term, the SNR of $^1$H imaging is usually higher than $\Cs$ imaging given high spin density and high gyromagnetic ratio. For the second term $\frac{s_{o}-s_{b}}{s_{o}}$ , our approach of direct $\Cs$ imaging can have much larger value than relaxation contrast imaging. In our approach, $\frac{s_{o}-s_{b}}{s_{o}} =  \frac{\eta n-n_b}{\eta n}$, where $n$ and $n_b$ are the nuclear spin concentration for the object and the background respectively, and $\eta \approx 300$ is the hyperpolarization enhancement factor. The local spin density where diamond particles situate is much higher than the 1\% $\Cs$ background, so we can estimate $\frac{n}{n_b} \leq 1$, resulting in $\frac{s_{o}-s_{b}}{s_{o}} \approx 1$. On the other hand, in the case of relaxation contrast imaging, we take the saturation recovery T$_1$ weighted imaging as an example and  $\frac{s_{o}-s_{b}}{s_{o}} = \frac{e^{-\mathrm{TR}/T_1} - e^{-\mathrm{TR}/T_{1b}}}{e^{-\mathrm{TR}/T_1}}$ , where $\mathrm{TR}$ is the echo time and $T_1$ and $T_{1b}$ are relaxation time for the object and background. Assuming $T_1$ = 0.5 ms and $T_{1b}$ = 0.1 ms, while $\mathrm{TR}$ = 5 ms, and this term $\sim$ 0.1, which is smaller than 1 in the case of hyperpolarized $\Cs$ imaging.

Combining the two terms in this example, we can conclude that in situations where the SNR of the 1H contrast imaging is less than 10 times of the $\Cs$ direct imaging, our approach presents a higher CNR. Similar comparison principle can be extended to more general cases: when the second term in the expression can offset the loss in the SNR term, our approach is better than $^1$H relaxation-based imaging approaches.

\end{document}

%% file: Commands3.tex
\newcommand{\NV}{NV$^{\text{-}}$\:}

\newcommand{\xd}{\delta}

\newcommand{\xt}{\theta}

\newcommand{\xl}{\lambda}
\newcommand{\xr}{\rho}

\newcommand{\app}{\approx}

\newcommand{\Cs}{{}^{13}\R{C}}

\newcommand{\Hs}{{}^{1}\R{H}}

\newcommand{\mI}[0]{\mathcal{I}}

\newcommand{\sinc}{\R{sinc}}

\newcommand{\xD}{\Delta}

\newcommand{\xO}{\Omega}

\newcommand{\fr}[2]{\frac{#1}{#2}}
\newcommand{\ov}[1]{\overline{#1}}

\newcommand{\mF}[0]{\mathcal{F}}

\newcommand{\rt}{\rightarrow}

\newcommand{\beq}{\begin{equation}}
\newcommand{\eeq}{\end{equation}}
                  
\newcommand{\benum}{\begin{enumerate}}
\newcommand{\eenum}{\end{enumerate}}
                    
\newcommand{\bit}{\begin{itemize}}
\newcommand{\eit}{\end{itemize}}

\newcommand{\bea}{\begin{eqnarray}}
\newcommand{\eea}{\end{eqnarray}}

\newcommand{\bev}{\begin{verbatim}}
\newcommand{\eev}{\end{verbatim}}

\newcommand{\zt}{\times}

\newcommand{\qt}{\tau}

\newcommand{\lb}{\left(}
\newcommand{\rb}{\right)}

\newcommand{\T}[1]{\textbf{#1}}
\newcommand{\I}[1]{\textit{#1}}
\newcommand{\R}[1]{\textrm{#1}}
\newcommand{\zfl}[1]{\protect\label{fig:#1}}
\newcommand{\zfr}[1]{Fig. \ref{fig:#1}}

\newcommand{\ket}[1]{\left\vert{#1}\right\rangle}

\newcommand{\expec}[1]{\left\langle #1\right\rangle}

\newcommand{\ba}{\left\{ \begin{array}{lr}}
\newcommand{\ea}{\end{array}\right.}

\newcommand{\BRd}[1]{\textcolor{red}{#1}} %RoyalBlue MidnightBlue

\newcommand{\blist}[1]{
 \begin{list}{#1}%$\ast\circ\bullet\Right
 \begin{align}
	 arrow
 \end{align}
 $\checkmark\star
  { \setlength{\itemsep}{3pt}
     \setlength{\parsep}{2pt}
     \setlength{\topsep}{3pt}
     \setlength{\partopsep}{0pt}
     \setlength{\leftmargin}{1em}
     \setlength{\labelwidth}{1em}
     \setlength{\labelsep}{0.5em} } }
\newcommand{\elist}{
  \end{list}  }

\DeclareMathSymbol{\vartheta}{\mathalpha}{letters}{"12}
\DeclareMathSymbol{\theta}{\mathalpha}{letters}{"23}
\DeclareMathSymbol{\phi}{\mathalpha}{letters}{"27}
\DeclareMathSymbol{\varphi}{\mathalpha}{letters}{"1E}

\newcommand{\bef}
{
\begin{figure}[htbp]
\centering
}

\newcommand{\eef}{\end{figure}}